\documentclass[fleqn,usenatbib]{mnras}

\usepackage{newtxtext,newtxmath}
\usepackage[T1]{fontenc}
\usepackage{ae,aecompl}

\usepackage{graphicx}	% Including figure files
\usepackage{amsmath}	% Advanced maths commands
\usepackage{amssymb}	% Extra maths symbols
\usepackage{subfigure}
\usepackage{multicol}

\title[Photoevaporation of proplyds]{Rapid destruction of protoplanetary discs due to external photoevaporation in star-forming regions}

\author[R. B. Nicholson et al.]{
Rhana B. Nicholson,$^{1,2}$\thanks{E-mail: r.b.nicholson@2011.ljmu.ac.uk (RBN)}
Richard J. Parker,$^{2}$\thanks{Royal Society Dorothy Hodgkin Fellow}
Ross P. Church,$^{3}$
Melvyn B. Davies,$^{3}$
\newauthor{ Niamh M. Fearon$^{4}$ and Sam R. J. Walton$^{1}$}
\\
$^{1}$Astrophysics Research Institute, Liverpool John Moores University, 146 Brownlow Hill, Liverpool, L3 5RF, UK\\
$^{2}$Department of Physics and Astronomy, The University of Sheffield, Hicks Building, Hounsfield Road, Sheffield, S3 7RH, UK\\
$^{3}$Lund Observatory, Department of Astronomy and Theoretical Physics, Box 43, SE-221 00, Lund, Sweden\\ 
$^{4}$School of Physics and Astronomy, The University of Manchester, Oxford Road, Manchester, M13 9PL, UK              
}

\date{Accepted XXX. Received YYY; in original form ZZZ}

\pubyear{2018}

\begin{document}
\label{firstpage}
\pagerange{\pageref{firstpage}--\pageref{lastpage}}
\maketitle

\begin{abstract}
We analyse $N$-body simulations of star-forming regions to investigate the effects of external far  and extreme ultra-violet photoevaporation from massive stars on protoplanetary discs. By varying the initial conditions of simulated star-forming regions, such as the spatial distribution, net bulk motion (virial ratio), and density, we investigate which parameters most affect the rate at which discs are dispersed due to external photoevaporation. We find that disc dispersal due to external photoevaporation is faster in highly substructured star-forming regions than in smooth and centrally concentrated regions.  Sub-virial star-forming regions undergoing collapse also show higher rates of disc dispersal than regions that are in virial equilibrium or are expanding. In moderately dense ($\sim$100\,M$_{\odot}$\,pc$^{-3}$) regions, half of all protoplanetary discs with radii $\geq$100\,AU are photoevaporated within 1 Myr, three times faster than is currently suggested by observational studies. Discs in lower-density star-forming regions ($\sim$10\,M$_{\odot}$\,pc$^{-3}$) survive for longer, but half are still dispersed on short timescales ($\sim$2\,Myr). This demonstrates that the initial conditions of the star forming regions will greatly impact the evolution and lifetime of protoplanetary discs. These results also imply that either gas giant planet formation is extremely rapid and occurs before the gas component of discs is evaporated, or gas giants only form in low-density star-forming regions where no massive stars are present to photoevaporate gas from protoplanetary discs.
\end{abstract}


%%%%%%%%%%%%%%%%%%%%%%%%%%%%%%%%%%%%%%%%%%%%%%%%%%

%%%%%%%%%%%%%%%%% BODY OF PAPER %%%%%%%%%%%%%%%%%%

\section{Introduction}
\label{introduction}
Protoplanetary discs (`Proplyds') are thin Keplerian discs around pre-main-sequence stars \citep{1994ApJ...429..781S} and are the birth places of planets. Proplyds form as a result of angular momentum conservation during the gravitational collapse of clouds when stars are forming. Within the discs, dust can coagulate to form a range of objects from pebbles to planets. 

The evolution and dispersal of protoplanetary discs controls the planet formation process, and observations suggest that disc lifetimes are between $\approx 3 - 5$ Myr \citep[e.g.][]{1995Natur.373..494Z,2001ApJ...553L.153H,2006ApJ...651.1177P,2018MNRAS.477.5191R}. Internal processes remove mass from the protoplanetary disc and, after several Myr, disc accretion slows significantly to the point that these processes begin removing more mass than can be replaced, leading to very rapid disc dispersal \citep{2001MNRAS.328..485C, 2011MNRAS.412...13O}.

Proplyd host stars do not form in isolation, but rather in clusters and associations with stellar densities that exceed that of the Galactic field by a few orders of magnitude \citep{2003ARA&A..41...57L,2010MNRAS.409L..54B}. Tens to thousands of stars can form in these regions that are a fraction of a parsec in size \citep{2000prpl.conf..151C}. Observations of young star-forming regions have revealed that stars form in filamentary structures \citep{2010A&A...518L.102A, 2014MNRAS.439.3275W}, resulting in hierarchical spatial distributions. The net motion of stars within these regions indicate that the structures are often collapsing (i.e. sub-virial) \citep{2006A&A...445..979P,2015ApJ...799..136F,2018arXiv180702115K}. 

The initial densities of star-forming regions are difficult to determine, and span a wide range \citep{2010MNRAS.409L..54B,2012MNRAS.421.2025K}, but many are thought to be at least $\sim$ 100 M$_{\odot}$ pc$^{-3}$ at the epoch of star formation \citep{2014MNRAS.445.4037P}. \cite{2014MNRAS.445.4037P} shows that two regions with similar present-day densities at present times may have originally had very different initial densities because initially dense regions expand much faster than lower-density regions due to two-body relaxation. They compare the present-day stellar densities and amount of spatial substructure in seven star-forming regions, including the Orion Nebula Cluster and Upper Scorpius -- which both contain massive stars that could act as external photoionising sources -- to infer the likely range of initial stellar densities in each of these star-forming regions and all  are consistent with having an initial density in the range  10 -- 1000 M$_{\odot}$ pc$^3$.

External processes, such as close stellar interactions, can also cause proplyds to be truncated or destroyed, as well as disrupting the orbits of fledgling planets \citep{2000A&A...362..968A, 2001MNRAS.323..785B, 2001MNRAS.325..449S, 2006ApJ...641..504A, 2008A&A...488..191O, 2012MNRAS.419.2448P, 2014MNRAS.441.2094R, 2015A&A...577A.115V, 2016MNRAS.457..313P, 2018arXiv180400013W}. The density of the star forming region will affect the rate of stellar interactions, with stars in low-density environments experiencing fewer dynamical interactions than higher density environments \citep{2014MNRAS.438..639W, 2010MNRAS.409L..54B}.  Furthermore, star-forming regions can contain massive stars (> 15\,M$_{\odot}$), whose intense far ultra-violet (FUV) and extreme ultra-violet (EUV) radiation fields are significantly higher than those in the interstellar medium \citep{2000A&A...362..968A,2004ApJ...611..360A, 2008ApJ...675.1361F}. This high-energy radiation heats the gaseous material of the upper layers of the disc until the thermal energy of the heated layer exceeds the gravitational potential of the disc, causing it to escape as a photoevaporative wind \citep{1994ApJ...428..654H,1998ApJ...499..758J}. This mass loss will affect the evolution of protoplanetary discs, and reduce the reservoir of material available to form gas giant planets \citep{2018MNRAS.475.5460H}. 

The effects of external photoevaporation appear to be observed in nearby star-forming regions, such as the Orion Nebula Cluster (ONC) \citep{1996AJ....111.1977M, 2016ApJ...826...16E, 2018ApJ...860...77E} and $\sigma$ Orionis \citep{2017AJ....153..240A}. The ONC has been preferentially observed due to its proximity to Earth ($\sim$415 pc) and because the discs can be viewed in silhouette due to the bright nebulous background. Studies have also shown that 80 -- 85\% of stars within the ONC host protoplanetary discs \citep{2000AJ....119.2919B, 2000AJ....120.3162L}, making the ONC a favorable target for studying disc evolution and dispersal \citep[though see][for an alternative interpretation which posits that the ONC proplyds are merely ionisation fronts of material left over from discs that are almost destroyed]{2007MNRAS.376.1350C}.

Due to two-body and violent relaxation, initially highly substructured star-forming regions can evolve to smooth and centrally concentrated clusters after only a few Myr \citep{2010MNRAS.407.1098A, 2014MNRAS.438..620P}. Furthermore,  two clusters that presently have similar densities may have had very different initial densities because initially very dense clusters expand faster than lower density counterparts. As mentioned before, the initial density will affect the rate at which protoplanetary discs are disrupted and destroyed due to stellar interactions. However, how much the initial density and substructure of a star-forming region affects the rate of protoplanetary disc dispersal due to external photoevaporation has yet to be studied.

Previous studies into the effects of external photoevaporation on protoplanetary discs in star-forming regions have tended to calculate the background UV radiation without directly calculating the disc mass-loss \citep{2000A&A...362..968A, 2004ApJ...611..360A}. \cite{2001MNRAS.325..449S} did calculate mass-loss rates in simulations specifically tailored to match the ONC, but assumed rather low stellar densities ($\sim$40\,M$_\odot$\,pc$^{-3}$), whereas \citet{2014MNRAS.445.4037P} suggests that the initial density of the ONC may have been much higher (>100\,M$_\odot$\,pc$^{-3}$).

These previous studies of external photoevaporation used spherically smooth spatial distributions with primordial mass segregation to model the environment of the ONC as observed today. However, observations of star forming regions show that stars form in highly substructured filamentary environments, where the stars are moving with subvirial velocities. The initial net motion and spatial structure of a star-forming region will govern its future evolution, and by extension the degree to which planet formation is hindered. These initial conditions lead to dynamical mass segregation on timescales of the age of the ONC \citep{2010MNRAS.407.1098A,2014MNRAS.438..620P}, negating the requirement for primordial mass segregation \citep{1998MNRAS.295..691B}.

Here, we focus on initial conditions that more closely reflect observations of young star forming regions \citep{2004MNRAS.348..589C}, and determine how much external photoevaporation affects the evolution of protoplanetary discs.. Therefore, we do not centrally concentrate our massive stars, but randomly distribute them in our simulated star-forming regions. We run suites of simulations that cover a range of initial conditions, with varying initial density, spatial distribution and net bulk motion (virial ratio). We then calculate and compare the mass-loss rates due to external photoevaporation for each set of initial conditions. 

The paper is organised as follows. In Section~\ref{method} we describe our $N$-body simulations, protoplanetary disc assumptions and our external photoevaporation prescription; in Section~\ref{results} we present our results; we provide a discussion in Section~\ref{discussion} and we conclude in Section~\ref{conclusion}.

\section{Method}
\label{method}
In this section we describe our method to select low-mass star-forming regions containing massive stars, before describing the subsequent $N$-body and stellar evolution of these regions.   

\subsection{Creating low-mass star-forming regions}
%%%RJP: DONT NEED THESE PARAGRAPHS IF WE'RE ONLY CHOSING TWO DIFFERENT MASSES (SEE LATER)
%To create our star-forming regions we first randomly sample masses from the following analytic power-law fit to the observed star cluster mass function (CMF):

%\begin{equation}
%N(M) \propto M^{-\beta_C},
%\end{equation}
%where $M$ is the cluster mass and $\beta_C$ describes the slope of the observed CMF. We adopt $\beta_C = 2$ \citep{1994A&AS..104..379B, 2003ARA&A..41...57L} and we sample this function for masses between $50 - 10^5$\,M$_\odot$.

We adopt two different masses ($100$ or $1000$M$_{\odot}$) for our star-forming regions and populate these regions with stars drawn randomly from the initial mass function (IMF) parameterised in \citet{2013MNRAS.429.1725M}, which has a probability density function of the form: 

\begin{equation}
p(m) \propto \left(\frac{m}{\mu}\right)^{-\alpha}\left(1 + \left(\frac{m}{\mu}\right)^{1 - \alpha}\right)^{-\beta}
\label{imf}.
\end{equation}
Here, $\mu = 0.2$\,M$_\odot$ is the average stellar mass, $\alpha = 2.3$ is the \citet{1955ApJ...121..161S} power-law exponent for higher mass stars, and $\beta = 1.4$ is used to describe the slope of the IMF for low-mass objects \citep*[which also deviates from the log-normal form;][]{2010ARA&A..48..339B}. Finally, we sample from this IMF within the mass range $m_{\rm low} = 0.1$\,M$_\odot$ to $m_{\rm up} = 50$\,M$_\odot$. 

We use a ``soft-sampling'' technique for sampling the IMF \citep{2006ApJ...648..572E}, which implies that the only formal limit on the most massive star that can form is that of the upper limit of the IMF \citep{2007MNRAS.380.1271P}. From an ensemble of Monte Carlo simulations, we find that typically, a $1000$M$_{\odot}$ star-forming region will contain five massive stars (M$_{\star} > 15$M$_{\odot}$). 

Low-mass star-forming regions rarely contain massive stars; however, if the only limit on the mass of the star that can form is the total mass of the star-forming region itself then occasionally we would expect a low-mass star-forming region to contain one or more massive stars and such regions are observed \citep[e.g. $\gamma^2$ Vel, a low-mass region containing at least two massive stars,][]{2014A&A...563A..94J}. Note that we are not explicitly attempting to model the $\gamma^2$ Vel star-forming region, which harbours a dense (sub)cluster within a more diffuse region. Instead, we are pointing out that observational examples of low-mass star-forming regions such as this and $\sigma$~Orionis occasionally contain massive stars and their photoevaporative effects on discs in these low-mass star-forming regions could be important.

In order to demonstrate the importance of low-mass star-forming regions, let us consider the demographics of star-forming regions from randomly sampling the underlying probability distributions. The observed mass function of star-forming regions follows a power-law of the form 
\begin{equation}
  N \propto M_{\rm cl}^{-2}, 
  \label{cmf}
\end{equation}
where $N$ is the number of star-forming regions with mass $M_{\rm cl}$ \citep{2003ARA&A..41...57L}, which implies that there many more low-mass star-forming regions compared to high-mass regions. 

We follow the procedure in \citet{2007MNRAS.380.1271P} and \citet{2017MNRAS.464.4318N} and sample $1 \times 10^6$ star-forming regions in the mass range 50 -- $10^5$\,M$_\odot$. Of these, $\sim$1200 lie in the mass range 1000$\pm$10\,M$_\odot$, and $\sim$ 15\,000 lie in the mass range 100$\pm$10\,M$_\odot$.

We then randomly populate our star-forming regions with stars drawn from Eqn.~\ref{imf}, until the total mass of stars equals or exceeds the chosen star-forming region mass from Eqn.~\ref{cmf}.

%From our sampling, the number of high mass (1000 Msun) clusters containing 3 or more massive stars is Ncluster = 577 (the total number containing any massive star is 996). The number of low mass (100 Msun) clusters containing 1 or 2 massive stars is Ncluster = 3122. The average number of stars in low mass clusters is 114, and the average number of stars in a high mass cluster is 1715. Therefore, the total number of stars from high mass clusters is 989,555 and the total from low mass clusters is 355,908; approximately 36% of the number in high mass clusters. This means that the number of stars being produced in low mass clusters containing massive stars is not at all insignificant.

%996 = 1000
%114 = 110
%1715 = 1710
%355908 = 341000
%170814 = 1700000
%577 = 580
%989555 = 986000
%36 = 35

\cite{2017MNRAS.464.4318N} found that $\sim$ 10 per cent of low-mass star-forming regions contain at least one massive star ($>$15\,M$_\odot$) when using the ``soft-sampling'' technique described above, and 1\,per cent of low-mass regions contain two massive stars. Furthermore, when taking into account the decreasing probability of forming a high-mass star-forming region (Eqn.~\ref{cmf}), the number of low-mass ($M_{\rm cl} = 100 \pm 1$\,M$_\odot$) regions containing at least one massive star is $\sim$3100, which is actually greater than the total number of high-mass ($M_{\rm cl} = 1000 \pm 10$\,M$_\odot$) star-forming regions (1200). Of these 1200 high-mass regions, $\sim$1000 contain at least one massive ($>$15\,M$_\odot$) star.

If we translate these numbers into the total number of stars that may be affected by photoevaporation, the average number of stars in our $M_{\rm cl} = 100 \pm 1$\,M$_\odot$ star-forming regions containing at least one massive star is $\sim$110, so in the 3100 low-mass regions that contain at least one massive star there are $\sim$341\,000 stars in total. The average number of stars in our high mass star-forming regions ($M_{\rm cl} = 1000 \pm 10$\,M$_\odot$) is $\sim$1710, and so the 1000 regions that contain at least one massive star host a total of $\sim$1\,700\,000 stars that could be affected by photoevaporation. In short, the fraction of \emph{stars} originating in low-mass star-forming regions containing at least one massive star is $\sim$20\,per cent of the total number of stars originating from high-mass regions containing at least one massive star. If we stipulate that the high-mass regions must contain three or more massive stars, only $\sim$580 regions out of 1200 fulfil this criteria and host a total of 986\,000 stars. The fraction of stars originating in low-mass star-forming regions containing at least one massive star is $\sim$35\,per cent of the total number of stars originating from high-mass regions containing at least three massive stars.

Crucially, this makes no assumption about the disruption and dissolution of these star-forming regions, and how many stars from each type of region eventually enter the Galactic field. The Galactic potential will influence the destruction of low-mass star-forming regions much more than high-mass regions \citep{2008gady.book.....B}, which take longer to dissolve into the Galactic field (and some remain as long-lived open clusters). Therefore, the majority of planet-hosting Field stars may come from lower-mass regions. 

%compared to  Furthermore, \citet{2007MNRAS.380.1271P} and \citet{2017MNRAS.464.4318N} find the interesting result that the \textbf{total} number of stars forming in low-mass star-forming regions ($M_{\rm cl} = 100$\,M$_\odot$) that contain one or two massive stars are expected to be as common an outcome of the star formation process as stars from more massive regions ($M_{\rm cl} = 1000$\,M$_\odot$), \textbf{with a ratio of 10:1 stars originating from low mass clusters with 1 massive star in comparison to 2.}

Given their significant contribution to the integrated stellar mass function, we therefore also investigate low-mass star-forming regions ($100$\,M$_{\odot}$) that contain either one or two massive stars -- these represent an unusual sampling of the IMF but allow us to investigate the effects of photoevaporation in less populous star-forming regions. %Low-mass star-forming regions containing several massive stars are observed \citep[e.g. $\gamma^2$ Vel,][]{2014A&A...563A..94J}, \textbf{are as common as massive clusters containing massive stars, when using this ``soft-sampling'' technique \citep{2017MNRAS.464.4318N} and may represent optimal initial conditions for the early Solar System.}

Hence we have three different star-forming region set-ups; a 100\,M$_{\odot}$ region with one massive star (38\,M$_\odot$), a 100\,M$_{\odot}$ region with two massive stars (42 and 23\,M$_\odot$) and one 1000 M$_{\odot}$ region with 5 massive stars (43, 33, 26, 17 and 17\,M$_{\odot}$). These regions were selected as the median outcomes of Monte Carlo sampling of $1 \times 10^6$ star-forming regions \citep{2017MNRAS.464.4318N}, and then filled with stellar masses drawn from the IMF \citep{2013MNRAS.429.1725M}. We then selected the median regions in terms of the total number of stars from within the mass ranges of 100\,$\pm1$M$_\odot$ and 1000\,$\pm10$M$_\odot$, with the stipulation that they had to contain massive stars. For the 100\,M$_\odot$ mass regions, we specifically selected the median region  containing one and two massive stars. For the 1000\,M$_\odot$ we selected the average cluster that contained three or more massive stars.

%These regions were selected as the median outcomes (when taking into account the mass of the cluster and the number of cluster members) of Monte Carlo sampling of 10\,000 star-forming regions \citep{2017MNRAS.464.4318N} of either 100\,M$_\odot$ or 1000\,M$_\odot$.

The external photoevaporation prescriptions we will adopt in this work are those from \citet{2001MNRAS.325..449S} which only weakly depend on the adopted stellar IMF, but in a future paper we will assign a FUV flux and an EUV flux to each intermediate/high-mass star based on its mass and then determine the respective fluxes incident on every low-mass star and use  the recent FRIED grid of models \citep{2018MNRAS.481..452H} to determine mass-loss for individual discs.

%\textbf{From the clusters that are generated, we select the average mass clusters that contain massive stars (> 15 M$_{\odot}$). To achieve this, we select the `median' 100 and 1000 M$_{\odot}$ of these clusters that contain massive stars. This results in three clusters, two 100 and one 1000 M$_{\odot}$ pc$^{-3}$, with 1, 2 and 5 massive stars (> 15 M$_{\odot}$) respectively. We specifically select these clusters containing massive stars so that the effects of external photoevaporation can be studied.}

\subsection{$N$-body simulations}
Our simulations are created with initial substructure by following the box-fractal method in \cite{2004A&A...413..929G}. We use a range of fractal dimensions for varying amounts of substructure: $D = 1.6$ (highly sub-structured), $D = 2.0$ (moderately sub-structured), and $D = 3.0$ (smooth). The method also correlates stellar velocities on local scales so that nearby stars have similar velocities, but more distant stars can have a wide range of different velocities.

The velocities are then scaled to the virial ratio $\alpha_{\rm vir}$, where $\alpha_{\rm vir} = T/|\Omega|$; $T$ is the total kinetic energy and $\Omega$ is the total potential energy of the stars. A range of virial ratios are used: $\alpha_{\rm vir} = 0.3$ (sub-virial, or collapsing), $\alpha_{\rm vir} = 0.5$ (virial equilibrium), and $\alpha_{\rm vir} = 0.7$ (supervirial, or expanding). Note that this virial ratio determines the net bulk motion, i.e.\,\,whether the star-forming region will collapse or expand. The correlated velocities on local scales mean that the local velocity dispersion can be subvirial, facilitating a violent relaxation process as the star-forming region evolves. Such local subvirial velocity dispersions are observed in the earliest stages of star formation \citep{1981MNRAS.194..809L,2006A&A...445..979P, 2015ApJ...799..136F}.

We create star-forming regions with stellar densities of 100\,M$_{\odot}$\,pc$^{-3}$ or  10\,M$_{\odot}$\,pc$^{-3}$ for the 1000\,M$_\odot$ regions; for the 100\,M$_\odot$ regions we set an initial density of 100\,M$_{\odot}$\,pc$^{-3}$. Such densities bracket the range observed in present-day star-forming regions \citep{2010MNRAS.409L..54B} as well as allowing for potentially higher \emph{primordial} densities \citep{2014MNRAS.445.4037P}.

Finally, we created a set of simulations with a Plummer sphere distribution \citep{1911MNRAS..71..460P} to facilitate comparisons with previous studies. We use the same IMF from our 1000 M$_{\odot}$ simulations to create two clusters with Plummer sphere distributions that have initial densities of 10 and 100 M$_{\odot}$ pc$^{-3}$.

We evolve each of our star-forming regions for 10 Myr using the \texttt{kira} integrator within the \texttt{Starlab} environment \citep{2001MNRAS.321..199P}. Stellar evolution is implemented using the \texttt{SeBa} look-up tables \citep{1996A&A...309..179P}. No binary or multiple stellar systems are included in these simulations. To gauge the amount of stochasticity in the disc photoevaporation, we run 20 realisations of the same initial conditions, identical apart from the random number seed used to assign the positions and velocities.

%The initial densities of the clusters are fixed; a moderate initial density which is consistent with observations of star forming regions ($100 M_{\odot} pc^{-3}$), and a low initial density ($10 M_{\odot} pc^{-3}$). By fixing the density, clusters with different initial virial rations and 

\subsection{Protoplanetary discs and external photoevaporation}

The mass loss rate of discs at a certain distance from a neighboring massive star is determined by the strength of the star's FUV ($h\nu$ < 13.6 eV) and EUV ($h\nu$ > 13.6 eV) fluxes at that distance. Mass loss due to FUV photons is caused  by heating the circumstellar disc, which creates an unbound neutral layer that can drift towards the ionization front, where it meets the EUV field. FUV is independent of the distance from the massive star because the only requirement is that the FUV flux is strong enough to heat the disc above its escape velocity. With EUV, the mass loss rate depends on the EUV flux and so is directly dependent on the distance from the massive star(s).

We use the same prescriptions for FUV and EUV photoevaporation as \cite{2001MNRAS.325..449S}:

\begin{equation}
\dot{M}_{\rm FUV} \approx 2 \times 10^{-9}r_{d} \,\,{\rm M_{\odot} yr^{-1}},
\end{equation}

\begin{equation}
\dot{M}_{\rm EUV} \approx 8 \times 10^{-12}r^{3/2}_{d}\frac{\Phi_{i}}{d^2} \,\,{\rm M_{\odot} yr^{-1}},
\end{equation}
where $r_{d}$ is the radius of the disc in astronomical units, $\Phi_{i}$ is the ionizing EUV photon luminosity from each massive star in units of 10$^{49}$ s$^{-1}$ and $d$ is the distance from the massive star in parsecs. The UV photon rate ($\Phi_{i}$) for the massive stars (> 15M$_{\odot}$) is dependent on stellar mass and we use the values from \citet{1996ApJ...460..914V} and \citet{2003ApJ...599.1333S}.

These photoevaporation rates were derived assuming a disc density profile $\Sigma \propto r_d^{-2}$ \citep{2000prpl.conf..401H,2009apsf.book.....H}; however, our analysis does not take into account the evolution of the surface density profile if the disc radius were to change significantly. 

%\textbf{One radius is selected for every protoplanetary disc in the cluster, and remains constant for the duration of the simulation.}

Observations of star forming regions show that disc radii can extend to several 100s AU \citep[e.g.][]{2018arXiv180305923A}. However, the typical initial radius of proplyds is still debated in the literature and therefore we sample a wide range of initial radii: 10, 50, 100, 200 and 1000 AU. We adopt a single value for each disc radius, focusing primarily on 100 AU discs, and then repeat the analysis for the five values. In reality, the radius of the disc will change due to internal processes such as viscous evolution, and due to internal and external photoevaporation, but we are unable to account for this (and the changing disc density profile) in our $N$-body simulations.

The initial disc masses are also debated, with theoretical constraints from the Minimum Mass Solar Nebula \citep[MMSN,][]{1977Ap&SS..51..153W, 1981PThPS..70...35H} and observations \citep{2013ApJ...771..129A,2016ApJ...828...46A} suggesting an upper limit of $M_{\rm disc} = 0.01$\,M$_{\star}$. Following the example in \cite{2001MNRAS.325..449S}, the initial disc masses in our simulation are 10 per cent of the host star mass ($M_{\rm disc}$ = 0.1 M$_{\star}$). Current observations suggest that disc masses are $\sim$1 per cent of the host star mass. We select 10 per cent so that we are sampling the upper range of the disc masses. While we do not account for accretion onto the protoplanetary discs, our discs are large enough in mass that we can neglect the accretion onto the disc as it will be minimal in comparison. For completeness, we ran a set of simulations where the disc masses were 1 per cent of the stellar host mass, which is more consistent with the MMSN estimates.

We subtract mass from our discs according to Eqns.~2~and~3. The rate of photoevaporation due to EUV radiation is dependent on distance from the ionising source, $d$, whereas the photoevaporation rate due to FUV is largely independent of distance from the source \citep{1999ApJ...515..669S}. Following \citet{1999ApJ...515..669S} and \citet{2001MNRAS.325..449S} we apply Eqn.~2 if the disc-hosting stars are within 0.3\,pc of the ionising source, noting that this distance is calibrated to models where $\theta^1$C~Ori is the most massive star (40\,M$_\odot$), which is commensurate with the most massive star in our simulations. However, we note that this is likely an underestimate of the amount of photoevaporation due to FUV fields in star-forming regions \citep{2016MNRAS.457.3593F,2018MNRAS.475.5460H}.

\section{Results}
\label{results}
We focus on 1000 M$_{\odot}$ star-forming regions, which typically contain $\sim$5 massive stars (M$_{\star} > 15$M$_{\odot}$) that act as photoionising sources \citep[c.f.][]{2001MNRAS.325..449S}. Our 1000 M$_{\odot}$ cluster contains 5 massive stars; 43.2, 32.7, 25.7 and two 17 M$_{\odot}$ with $\Phi_{i}$ values of 1.1, 0.47, 0.19 and $\sim$ 0.013 respectively.  We focus on the results for two different initial stellar densities, $\sim$ 10 M$_{\odot}$ pc$^{-3}$ and $\sim$ 100 M$_{\odot}$ pc$^{-3}$ and, apart from the final section, the assumed initial mass for every disc is $M_{\rm disc} = 0.1$\,M$_{\star}$.

We present the results from varying different initial properties within the star forming regions, focusing on the spatial distribution (fractal dimension, $D$) and net bulk motion (virial ratio, $\alpha_{\rm vir}$). We focus on protoplanetary discs that have a radius of 100 AU, however the effects of external photoevaporation on discs with different radii and mass are discussed later in Sections~\ref{dispersal rates} and \ref{disc masses}.

We compare our results to the observed disc fractions in both \cite{2001ApJ...553L.153H} and \cite{2018MNRAS.477.5191R} using ages from the models in \cite{2000A&A...358..593S}. We discuss the caveats associated with these models in Section~4.

We later present two low mass clusters (100 M$_{\odot}$) with an initial density of $\sim$ 100 M$_{\odot}$ pc$^{-3}$ that are subvirial ($\alpha_{\rm vir} = 0.3$) and highly substructured ($D =$ 1.6). Our clusters contain either one (38\,M$_\odot$) or two (42\,M$_\odot$ and 23\,M$_\odot$) massive stars. The corresponding $\Phi$ values are $\Phi = 0.76$ and $\Phi = 1.01, 0.11$ respectively.

\subsection{Substructure in star-forming regions}
We first present the results from four simulations of star forming regions, where in each simulation the star forming region has a different initial spatial distribution; $D =$ 1.6 (highly substructured), $D = 2.0$ (moderately substructured), $D = 3.0$ (smooth) and a Plummer sphere spatial distribution. In all simulations the star-forming region is subvirial ($\alpha_{\rm vir} = 0.3$).

In Fig.~1 we show the average fraction of stars that have retained their (100 AU) discs from 20 runs of each simulation, where the initial substructure of the star-forming region is varied. We present the results for regions with two different initial stellar densities; 10 and 100 M$_{\odot}$ pc$^{-3}$ respectively. 

\begin{figure*}
	\begin{center}
    \setlength{\subfigcapskip}{10pt}
    \subfigure[Density = 10 M$_{\odot}$ pc$^{-3}$]{\label{high_1000_f}{\includegraphics[scale=0.433]{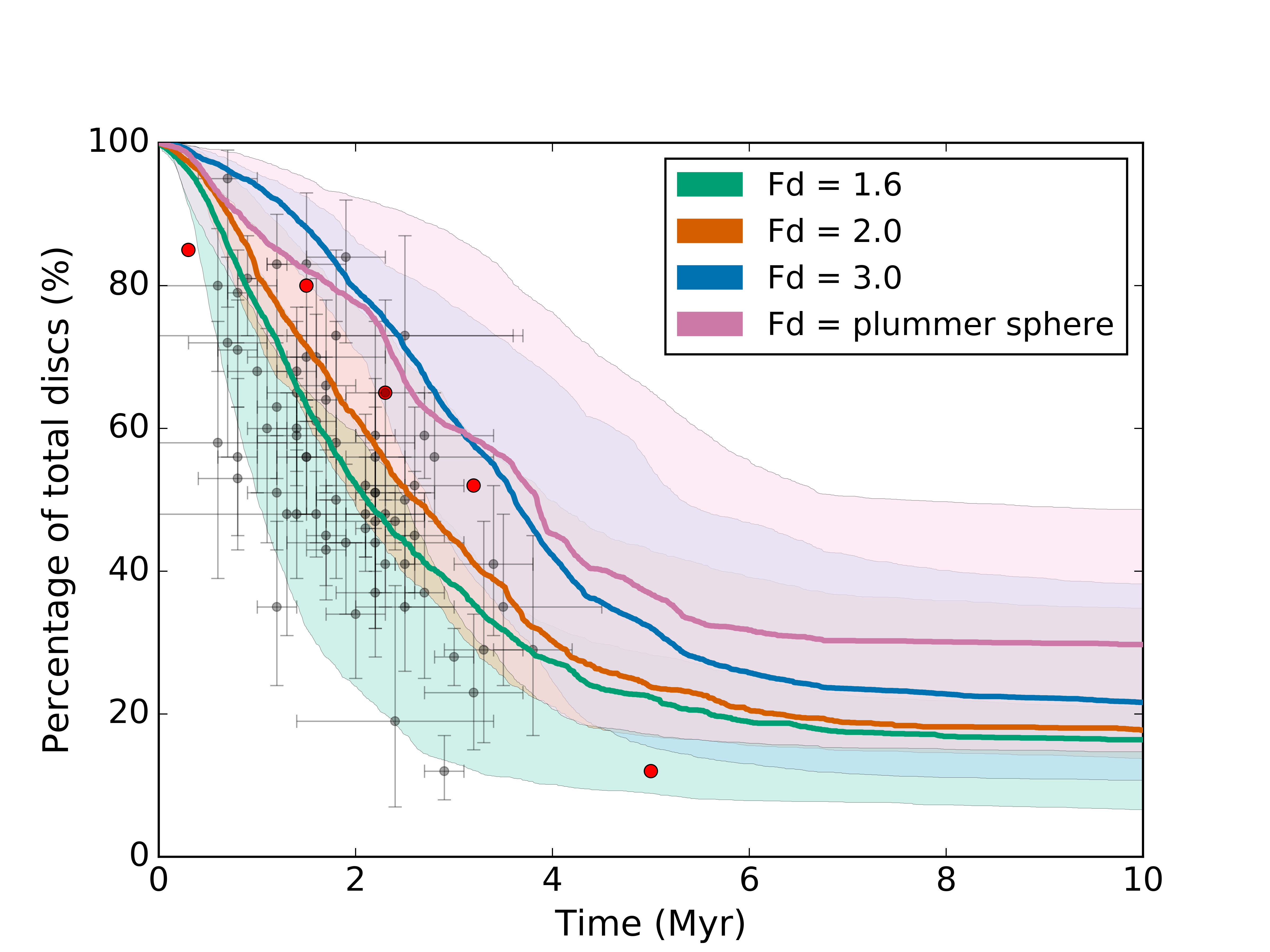}}}
    \subfigure[Density = 100 M$_{\odot}$ pc$^{-3}$]{\label{low_1000_f}{\includegraphics[scale=0.433]{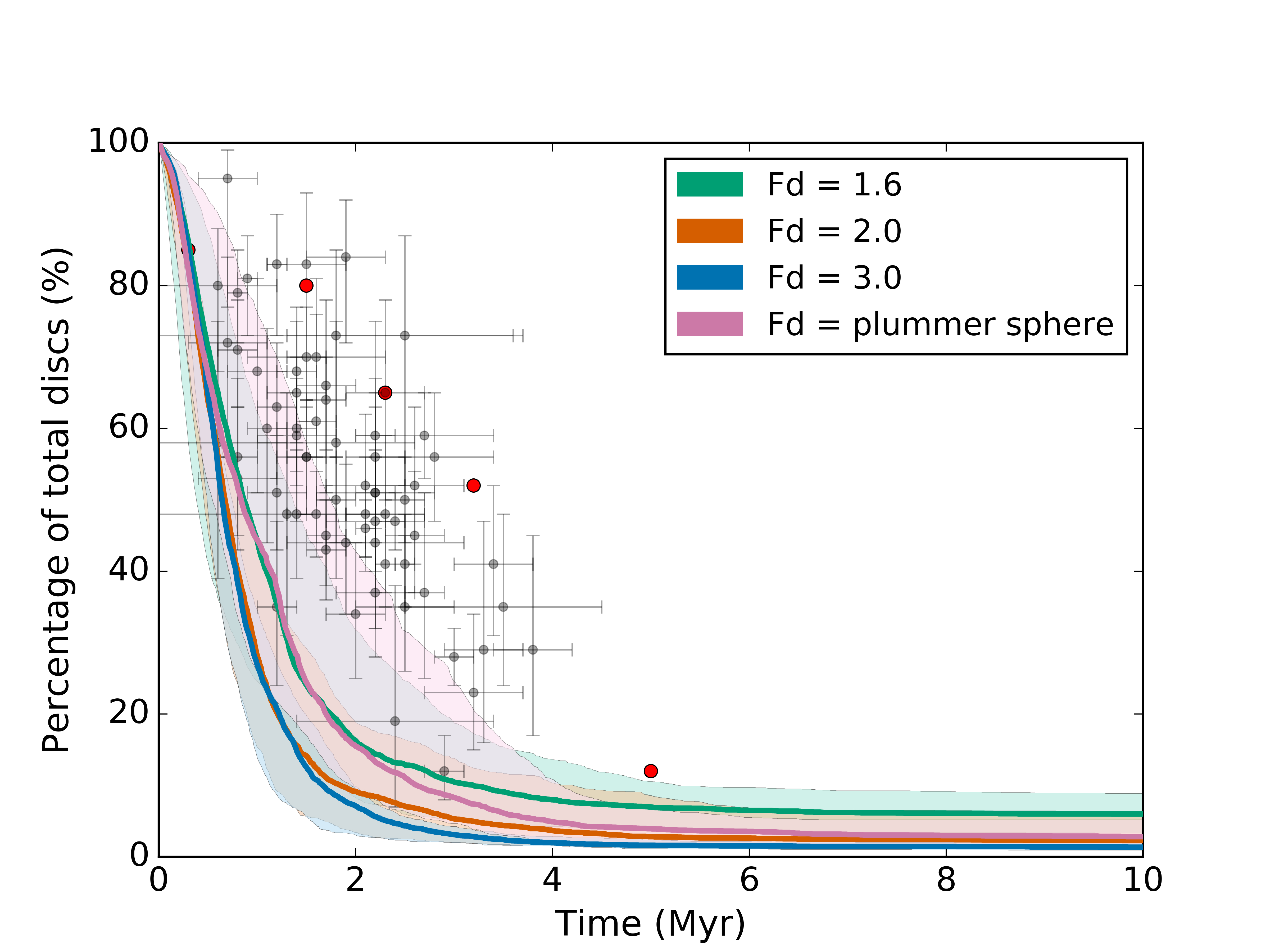}}}
    \caption{The average percentage of stars retaining their 100 AU disc over time within a sub-virial ($\alpha_{\rm vir} = 0.3$) cluster. The amount of substructure in the star-forming region is varied from highly substructured ($D = 1.6$) to smooth and centrally concentrated (Plummer Sphere). Two different initial densities (10 and 100 M$_{\odot}$ pc$^{-3}$) are considered. Each coloured line represents a different initial spatial distribution. The red data points are observational values from \protect\cite{2001ApJ...553L.153H}. The grey data points are from \protect\cite{2018MNRAS.477.5191R} using stellar ages from the models in \protect\cite{2000A&A...358..593S}. The colored shaded regions show the complete range of values from the 20 runs for each set of initial conditions.}
    \end{center}
\end{figure*}

Fig.~1(a) shows the results from a star forming region with an initial density of 10 M$_{\odot}$ pc$^{-3}$. Within highly sub-structured regions ($D = 1.6$), the time taken for half of the stars to lose their discs due to external photoevaporation is 2.12 Myr. In moderately sub-structured regions ($D = 2.0$), this time increases to 2.60 Myr. However,  the average percentage of remaining discs with time in both remain relatively similar throughout the 10 Myr. For regions with an initially smooth and spherical distribution ($D = 3.0$), the time taken for half of the discs to disperse is 3.62 Myr. Discs within Plummer spheres have the longest lifetimes (3.85\,Myr), with an average of $\sim$ 29.7 per cent of discs surviving for longer than 10 Myr. 

\citet{1911MNRAS..71..460P} models (and other models that describe smooth star clusters such as a \citet{1966AJ.....71...64K} profile or an \citet{1987ApJ...323...54E} profile) are intended the model dynamically relaxed systems, whereas young star-forming regions have yet to relax. Therefore, even a smooth box fractal ($D = 3.0$) contains kinematic substructure, which causes the dynamical evolution of such a region to be more violent than a smooth Plummer sphere. It is therefore unsurprising that fewer discs survive in kinematically substructured fractal regions than in Plummer spheres with a similar density. %\textbf{The plateaus within the Plummer Sphere results partially correspond with the deaths of the most massive stars (3.92, 4.70 and 5.82 Myr).}

Fig.~1(b) shows the results for star forming regions with an initial density of 100 M$_{\odot}$ pc$^{-3}$. For discs in the highly sub-structured regions ($D = 1.6$), the time taken for half of the stars within the cluster to lose their protoplanetary discs is 0.87 Myr. The majority of discs within the highly substructured region ($D = 1.6$) are dispersed after 10 Myr, with $\sim$ 6 per cent surviving for the length of the simulation. The majority of discs within smooth, spherical regions are also dispersed within a short time frame, with only $\sim$ 3 per cent remaining after 10 Myr. %All 4 of the different spatial distributions for these moderately dense clusters follow a similar trend with regard to disc survival rates.

In Table 1 we summarise the average time taken for half of the stars in a star forming region to lose their discs for each spatial distribution and Table 2 summarises the percentage of discs remaining at the end of the 10 Myr simulation.

In the low density simulations, regions with more spatial substructure photoevaporate discs faster than smoother regions of a comparable density (Fig.~1(a)) . The reason for this is that the more substructured regions are initially further from dynamical equilibrium than the smooth regions, and low-mass stars undergo more  close interactions with the high mass stars as the regions relax.

Interestingly, in the high-density simulations (Fig.~1(b)), whilst the fraction of discs remaining after 10\,Myr is lower than in the low density simulations,  the initially more substructured star-forming regions contain more discs than the smooth regions after 10\,Myr (and their disc-destruction half-life is longer, see Table 1). We attribute this to the higher ejection rate of massive stars in dense, substructured star-forming regions \citep{2014MNRAS.438..620P}, which means that some of the ionising sources are no longer near the majority of the proplyds as early as 1\,Myr after the start of dynamical evolution.

%The spread caused by different spatial distributions varies between low and moderately dense clusters. Within low density clusters, there is a noticeable difference between a highly substructured cluster and a cluster with a Plummer Sphere distribution. Within moderately dense clusters there is little to no difference.

\begin{table*}
	\centering
	\caption{The time taken for half of stars within the star-forming region to lose the gas from their 100 AU protoplanetary discs in a 1000 M$_{\odot}$ sub-virial ($\alpha_{\rm vir} = 0.3$) region with two different initial densities; 10 and 100 M$_{\odot}$ pc$^{-3}$. Four different spatial distributions are analysed; $D = 1.6$ (highly sub-structured), $D = 2.0$ (moderately sub-structured), $D = 3.0$ (smooth), and a Plummer sphere distribution. The highest and lowest values from the 20 different runs are included.}
	\begin{tabular}{ccccc} % four columns, alignment for each
		\hline
        & \multicolumn{2}{c|}{Half life of protoplanetary discs} \\
		Fractal dimension ($D$) & $\rho = $10 M$_{\odot}$ pc$^{-3}$ &  $\rho = $100 M$_{\odot}$ pc$^{-3}$\\
		\hline
		1.6 & 2.12 $\pm$ $^{0.51}_{1.11}$ Myr & 0.87 $\pm$ $^{0.50}_{0.49}$ Myr \\ \hline
        2.0 & 2.60 $\pm$ $^{1.36}_{0.62}$ Myr & 0.67 $\pm$ $^{0.21}_{0.22}$ Myr \\ \hline
        3.0 & 3.62 $\pm$ $^{1.68}_{0.89}$ Myr & 0.65 $\pm$ $^{0.10}_{0.16}$ Myr\\ \hline
        Plummer Sphere & 3.85 $\pm$ $^{3.70}_{1.34}$ Myr & 0.84 $\pm$ $^{0.90}_{0.29}$ Myr \\
		\hline
	\end{tabular}
\end{table*}

%More than half of the discs within both the smooth, centrally concentrated and Plummer Sphere regions survive the whole simulation run. %Table 2 summarises the percentage of discs remaining at the end of the 10 Myr simulation.

\begin{table*}
	\centering
	\caption{The average percentage of 100 AU discs remaining after 10 Myr within a sub-virial ($\alpha_{\rm vir} = 0.3$) star forming region from 20 realisations of each simulation. The amount of substructure is varied from highly substructured ($D = 1.6$) to a smooth and centrally concentrated Plummer sphere. The highest and lowest values from the 20 different runs are included. Two different initial densities (10 and 100 M$_{\odot}$ pc$^{-3}$) are considered.}
	\begin{tabular}{ccccc} % four columns, alignment for each
		\hline
        & \multicolumn{2}{c|}{Percentage of discs remaining after 10 Myr} \\
		Fractal dimension ($D$) & $\rho = $10 M$_{\odot}$ pc$^{-3}$ & $\rho$ = 100 M$_{\odot}$ pc$^{-3}$\\
		\hline
		1.6 &  16.40 $\pm$ $^{4.58}_{9.8}$ \% & 5.99 $\pm$ $^{2.88}_{2.67}$ \% \\ \hline
        2.0 & 17.75 $\pm$ $^{17.03}_{4.04}$ \% & 2.27 $\pm$ $^{4.04}_{0.92}$ \% \\ \hline
        3.0 & 21.60 $\pm$ $^{16.63}_{10.88}$ \% & 1.35 $\pm$ $^{0.63}_{0.42}$ \% \\ \hline
        Plummer Sphere & 29.70 $\pm$ $^{18.95}_{14.98}$ \% & 2.81 $\pm$ $^{2.36}_{1.63}$ \% \\
		\hline
	\end{tabular}
\end{table*}

%The percentage of discs surviving to 10 Myr decreases with increasing degrees of substructure. This effect is amplified in the moderately dense star forming region.

\subsection{Virial ratio}
We explore how changing the net bulk motion of the star-forming region affects the rate of disc dispersal due to external photoevaporation. We run simulations of our star-forming region with three different virial ratios; 0.3 (subvirial, or collapsing), 0.5 (virial equilibrium), and 0.7 (supervirial, or expanding). We keep the fractal dimension constant, adopting $D$ = 2.0 and as before we analyse  simulations with  two different initial densities; 10 M$_{\odot}$ pc$^{-3}$  and 100 M$_{\odot}$ pc$^{-3}$.

In Table 3 we summarise the average time taken for half of the stars in each region to lose their (100 AU) discs for a given bulk virial ratio. Table 4 shows the percentage of discs remaining at the end of the 10 Myr simulation. Fig.~2 shows the the average fraction of stars that have retained their discs from the 20 runs of each simulation for a star forming region where we vary the initial bulk motion (virial ratio).% with two different initial densities; 10 and 100 M$_{\odot}$ pc$^{-3}$ respectively.

\begin{figure*}
	\begin{center}
    \setlength{\subfigcapskip}{10pt}
    \subfigure[Density = 10 M$_{\odot}$ pc$^{-3}$]{\label{low_1000_f}{\includegraphics[scale=0.2775]{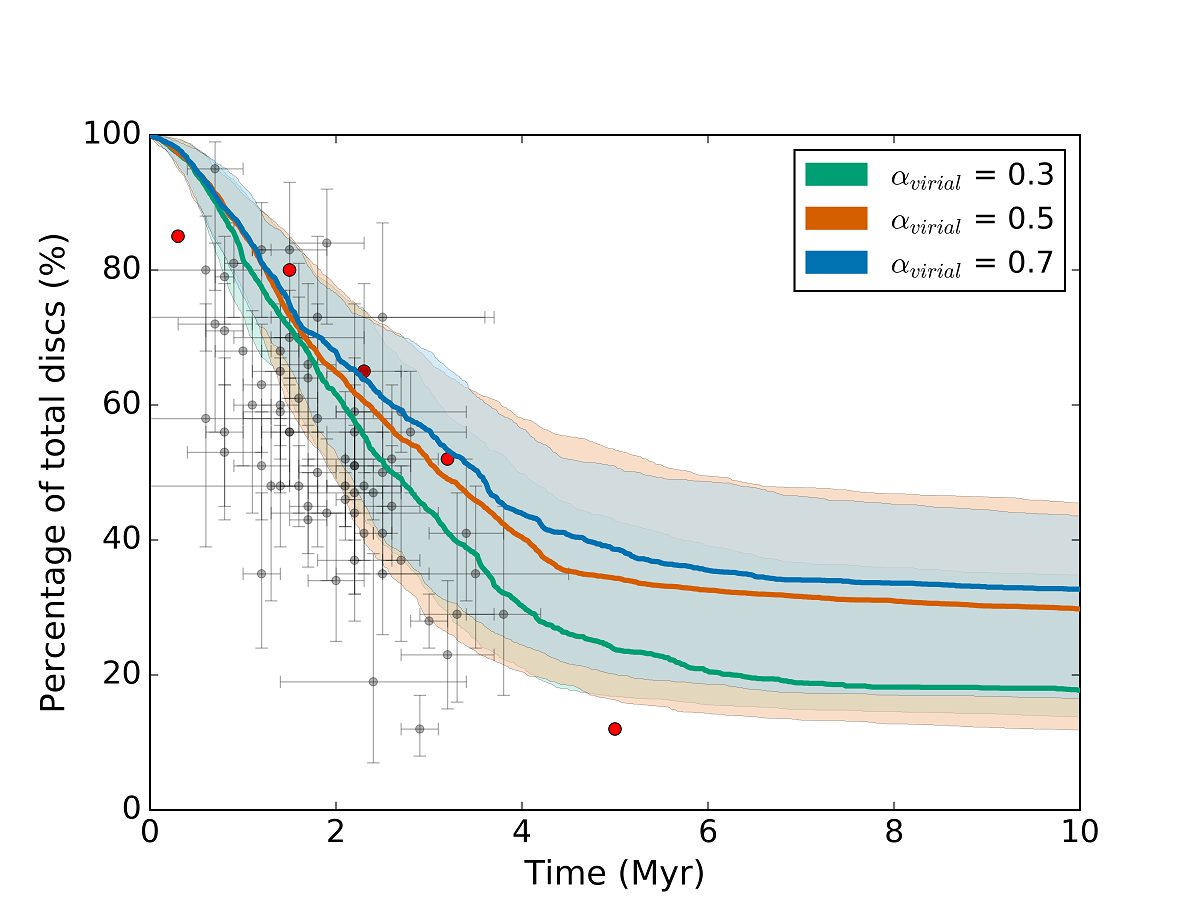}}}
    \subfigure[Density = 100 M$_{\odot}$ pc$^{-3}$]{\label{high_1000_f}{\includegraphics[scale=0.433]{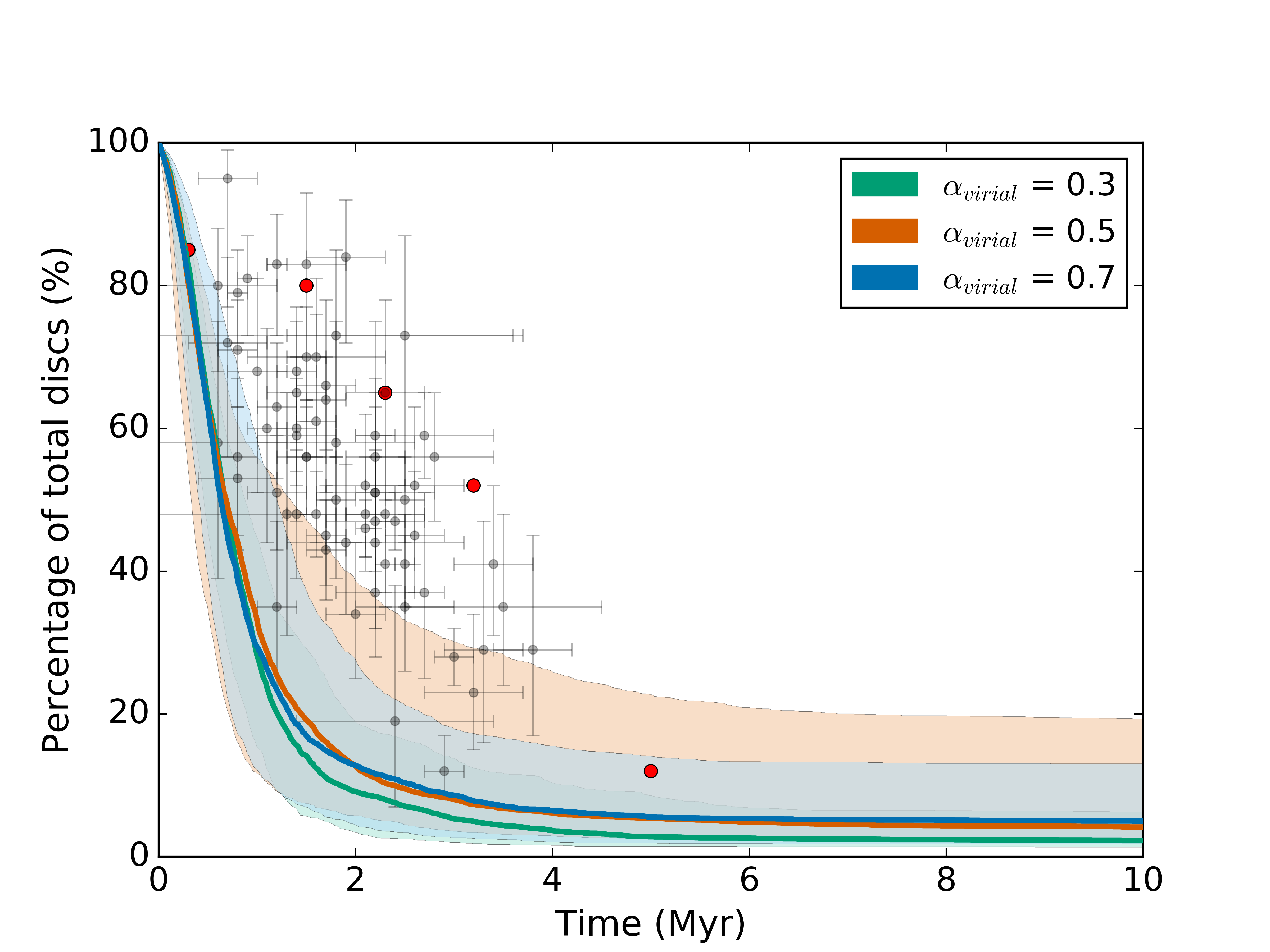}}}
    \caption{The average percentage of stars retaining their 100 AU disc with time in a 1000 M$_{\odot}$, moderately substructured ($D = 2.0$) star forming region with an initial density of 10 and 100 M$_{\odot}$ pc$^{-3}$. Each coloured line represents a different virial ratio. The red data points are observational values from \protect\cite{2001ApJ...553L.153H}. The grey data points are from \protect\cite{2018MNRAS.477.5191R} using ages from the stellar model in \protect\cite{2000A&A...358..593S}. The shaded regions show all values between the maximum and minimum values from all 20 runs of the simulations.}
    \end{center}
\end{figure*}

Fig.~2 (a) shows the average mass loss rate in a star forming region with an initial density of 10 M$_{\odot}$ pc$^{-3}$. The time taken for half of the stars within a collapsing (sub-virial) star-forming region to lose their discs is 2.60 Myr. In regions that are expanding (supervirial), this time increases to 3.53 Myr. The percentage of discs within the subvirial region after 10 Myr is 17.75\%, in comparison to discs within an expanding region where the percentage rises to 32.67\%. %All three different virial ratios have a very similar trend with regards to disc lifetime. 

The initial net bulk motion of low density star-forming regions affects the amount of discs that are photoevaporated due to external radiation, with subvirial regions evaporating more discs at a faster rate than either virialised or supervirial regions. 

Fig.~2(b) shows the results for a star-forming region with an initial density of 100 M$_{\odot}$ pc$^{-3}$. The time taken for half the stars within a collapsing region to lose their discs is 0.67 Myr. This time is similar for regions in virial equilibrium and expanding regions (0.68 and 0.63 Myr respectively). The lower disc half-life for the supervirial regions could again be due to massive stars being ejected in the (sub)virial regions.  The percentage of discs remaining after 10 Myr in sub-virial star forming regions is 2.27\%  whereas in regions where the net motion is expansive, this is increased to 5.00\%. %There seems to be little effect on disc dispersal rates by changing the virial ratio in these moderately dense star-forming regions.

%Lower case AU in fig 3 and 4 should be changed

\begin{table*}
	\centering
	\caption{The time taken for half of stars within the cluster to lose the gas within their 100 AU protoplanetary discs in a 1000 M$_{\odot}$, moderately sub-structured ($D = 2.0$) star forming region for two different initial densities; 10 and 100 M$_{\odot}$ pc$^{-3}$. Three different virial ratios are analysed: $\alpha_{\rm vir} = 0.3$ (sub-virial, or collapsing), $\alpha_{\rm vir} = 0.5$ (virial equilibrium), and $\alpha_{\rm vir} = 0.7$ (super virial, or expanding).}
	\begin{tabular}{ccccc} % four columns, alignment for each
		\hline
        & \multicolumn{2}{c|}{Half life of protoplanetary discs} \\
		Virial Ratio ($\alpha_{\rm vir}$) & $\rho = $10 M$_{\odot}$ pc$^{-3}$ &  $\rho = $100 M$_{\odot}$ pc$^{-3}$\\
		\hline
		0.3 & 2.60 $\pm$ $^{1.36}_{0.62}$ Myr & 0.67 $\pm$ $^{0.21}_{0.22}$ Myr \\ \hline
        0.5 & 3.10 $\pm$ $^{2.73}_{1.15}$ Myr & 0.68 $\pm$ $^{0.66}_{0.35}$ Myr \\ \hline
        0.7 & 3.53 $\pm$ $^{1.72}_{1.40}$ Myr & 0.63 $\pm$ $^{0.57}_{0.23}$ Myr\\
		\hline
	\end{tabular}
\end{table*}

\begin{table*}
	\centering
	\caption{The average percentage from 20 runs of simulations of 100 AU discs remaining after 10 Myr within a moderately substrutured ($D = 2.0$) cluster. The bulk motion (virial ratio) of the star-forming region is varied, from collapsing (sub-virial, $\alpha_{\rm vir} = 0.3$) to expanding (super virial, $\alpha_{\rm vir} = 0.7$). The highest and lowest values from the 20 different runs are included.}
	\begin{tabular}{ccccc} % four columns, alignment for each
		\hline
        & \multicolumn{2}{c|}{Percentage of discs remaining after 10 Myr} \\
		Virial Ratio ($\alpha_{\rm vir}$) & $\rho = $10 M$_{\odot}$ pc$^{-3}$ &  $\rho = $100 M$_{\odot}$ pc$^{-3}$\\
		\hline
		0.3 & 17.75 $\pm$ $^{17.03}_{4.04}$ \% &  2.27 $\pm$ $^{4.04}_{0.92}$ \% \\ \hline
        0.5 & 29.77 $\pm$ $^{15.73}_{17.95}$ \% &  4.16 $\pm$ $^{15.14}_{2.06}$ \% \\ \hline
        0.7 & 32.67 $\pm$ $^{10.9}_{16.14}$ \% &  5.00 $\pm$ $^{8.04}_{3.23}$ \% \\ 
		\hline
	\end{tabular}
\end{table*}

\subsection{Disc radii}
\label{dispersal rates}
Here we present the rates of disc dispersal for different initial disc radii in a star-forming region with two different initial densities (10 and 100 $M_{\odot}$ pc$^{-3}$). The region has a fractal dimension of $D = 2.0$ (moderately substructured) and a viral ratio of $\alpha_{\rm vir} = 0.3$ (sub-virial).

Fig.~3 shows the percentage of protoplanetary discs with initial radii ranging between 10 -- 1000\,AU that have some remaining mass over 10 Myr in a 1000 M$_{\odot}$ star-forming region with different initial stellar densities; 10 M$_{\odot}$ pc$^{-3}$ in Fig~3(a) and 100 M$_{\odot}$ pc$^{-3}$ in Fig.~3(b).

%Both figures show protoplanetary discs within the 1000 M$_{\odot}$ cluster with initial masses of 10 percent that of their host star with different disc radii. Figure 3(a) and (b) have different initial densities; 100 M$_{\odot}$ pc$^{-3}$ and 10 M$_{\odot}$ pc$^{-3}$ respectively. Figure 4 shows discs within the 1000 M$_{\odot}$ cluster with a density of 100 M$_{\odot}$ pc$^{-3}$ with initial disc masses that are 1 percent that of their host star.

\begin{figure*}
	\begin{center}
    \setlength{\subfigcapskip}{10pt}
    \subfigure[Density = 10 M$_{\odot}$ pc$^{-3}$]{\label{low_1000_f}{\includegraphics[scale=0.275]{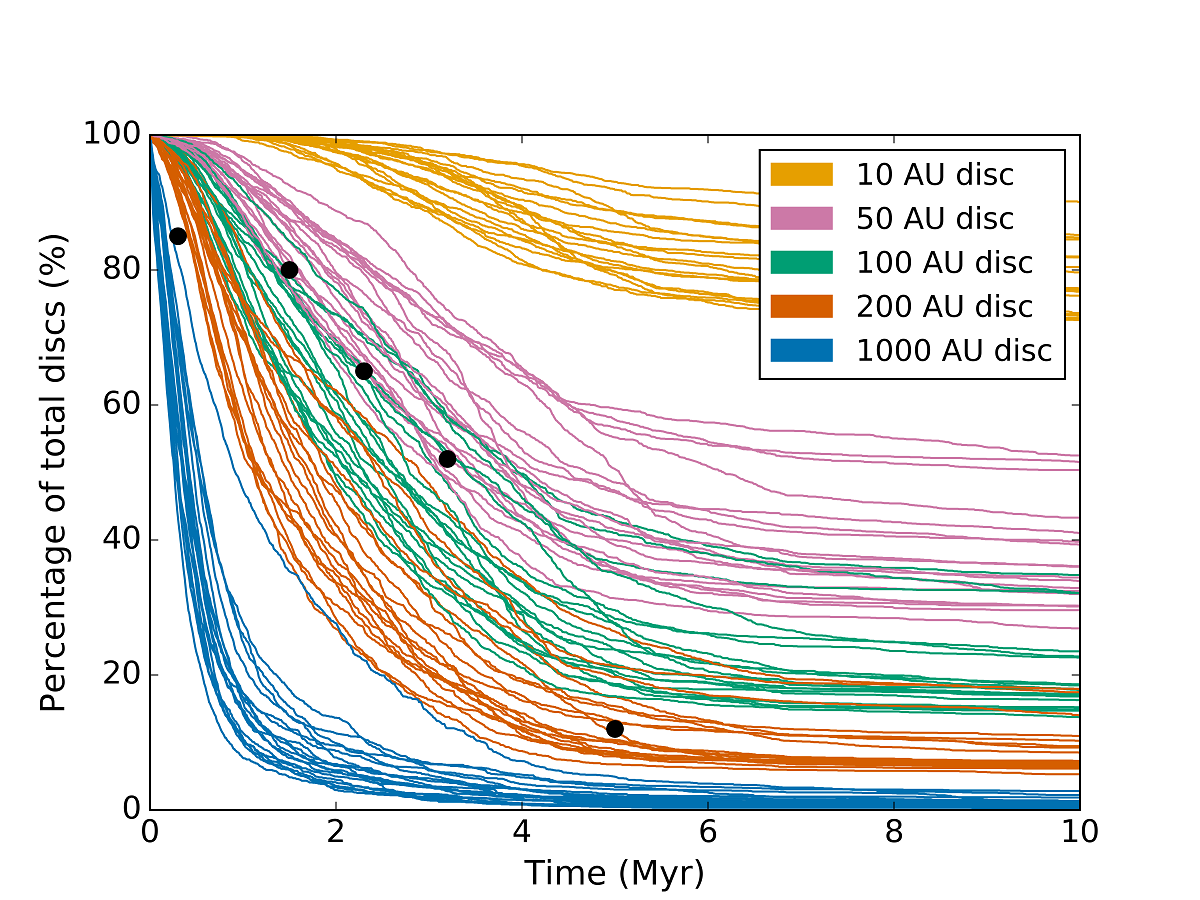}}}
        \subfigure[Density = 100 M$_{\odot}$ pc$^{-3}$]{\label{high_1000_f}{\includegraphics[scale=0.275]{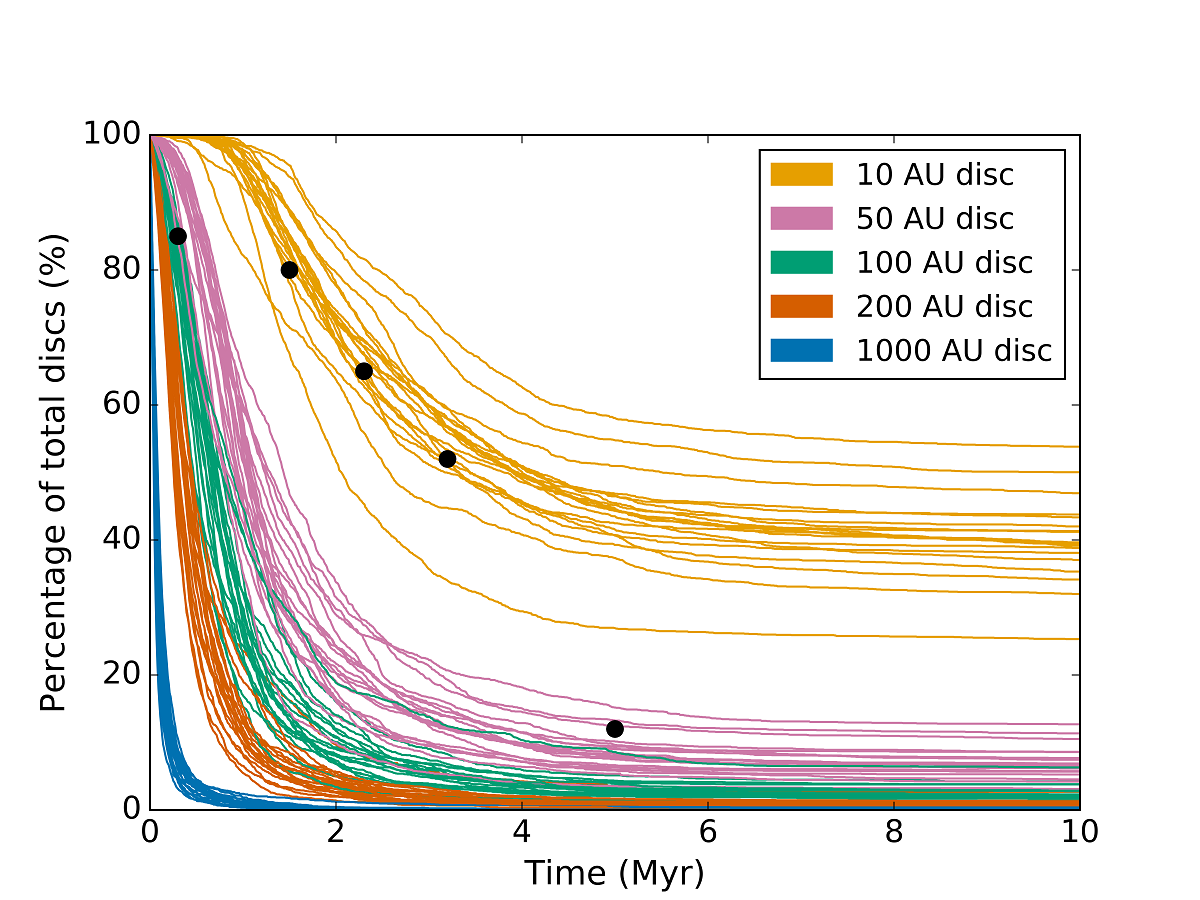}}}
    \caption{The percentage of total remaining discs over time for a 1000 M$_{\odot}$ star-forming region with an initial density of $\sim$10 and $100$ M$_{\odot}$ pc$^{-3}$ (panels a and b respectively). The cluster has is moderately sub-structured ($D = 2.0$) and is sub-virial ($\alpha_{vir} = 0.3$) and Each colour represents a different initial disc radius. The disc masses are 10 per cent of the host star mass. The multiple coloured lines are each a single run of the 20 simulation runs. The black data points are observational values from \protect\cite{2001ApJ...553L.153H}.}
    \end{center}
\end{figure*}

%We focus mainly on 100 Au discs, which is towards the lower end of the observed range of disc radii \citep{2015A&A...577A.115V}. We keep the disc masses at 10 percent of the initial mass of the star for all disc radii.

In the lower-density star-forming regions (Fig.~3(a)), the time taken for half of the 100 AU discs to be completely photoevaporated is 2.60 Myr. Discs with radii of 10 AU have much greater lifetimes, with an average of $\sim$77 per cent of discs surviving the full length of the simulation. The majority of discs with very large radii (1000 AU) are still depleted within very short timescales. Disc depletion rates begin to switch off after $\sim$4 Myr due to a combination of a large decrease in density of the star forming region, which peaks at $\sim$2 Myr, and the death of the most massive star at 4.33 Myr.

Fig.~3(b) shows that the majority (90\,per cent or more) of discs with radii $>$ 10\,AU are completely photoevaporated before the end of the 10 Myr simulation in moderately dense star-forming regions. The time taken for half of the stars in the region to lose their 100 AU discs is $0.67$ Myr. The vast majority of the largest discs (1000 AU) are photoevaporated completely within 2 Myr, with half of the stars in the region losing their discs within < 0.1 Myr.

%We later present our low mass cluster (100 M$_{\odot}$) with an initial density of $\sim$ 100 M$_{\odot}$ pc$^{-3}$. This cluster contains two massive stars (42\,M$_\odot$ and 23\,M$_\odot$).http://mnras.oxfordjournals.org/

We also ran simulations for low mass star-forming regions (100 M$_{\odot}$) with an initial density of $\sim$100 M$_{\odot}$ pc$^{-3}$. These low-mass regions contain two massive stars (42\,M$_\odot$ and 23\,M$_\odot$), which represents an unusual sampling of the IMF \citep{2007MNRAS.380.1271P}, but is observed in nature \citep[e.g. $\gamma^2$ Vel,][]{2014A&A...563A..94J}. In these low mass regions, half of discs with radii of 100 AU dissipated in 0.51 Myr. This is comparable to the 1000 M$_{\odot}$ regions with a similar density. We will further explore the effects of different stellar IMFs on disc dispersal in a future paper.

\subsection{Disc Masses}
\label{disc masses}

We have assumed that the disc masses are 10\,per cent of the host star's mass, which is likely to be an overestimate and various studies suggest that the disc mass is as low as 1\,per cent of the host star's mass \citep{1977Ap&SS..51..153W, 1981PThPS..70...35H, 2013ApJ...771..129A}.

In Fig.~4 we show the results for a star-forming region with our two different initial densities (10 and 100 M$_{\odot}$ pc$^{-3}$ respectively), where the initial disc masses are set to $M_{\rm disc} = 0.01$\,M$_{\star}$. 

\begin{figure*}
	\begin{center}
    \setlength{\subfigcapskip}{10pt}
    \subfigure[Density = 10 M$_{\odot}$ pc$^{-3}$]{\label{low_1000_f}{\includegraphics[scale=0.275]{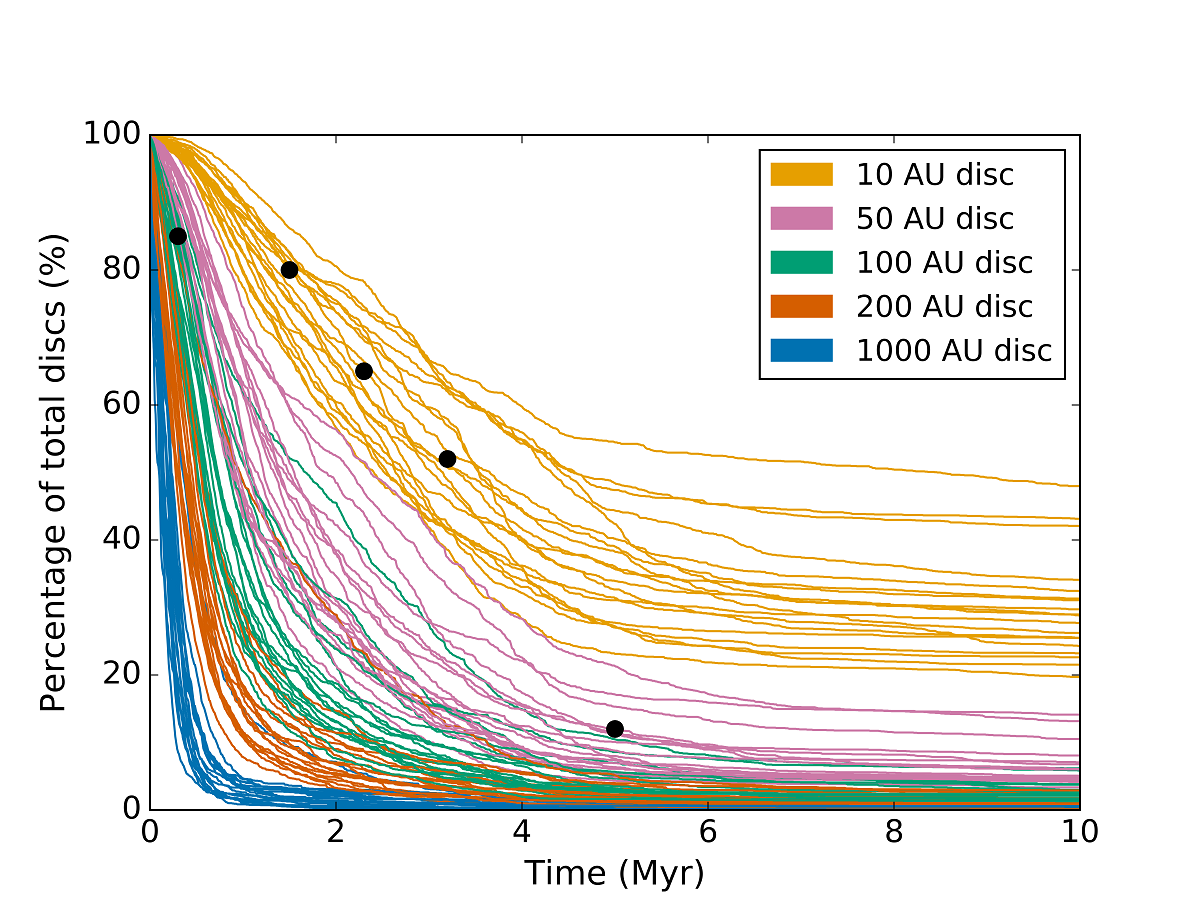}}}
    \subfigure[Density = 100 M$_{\odot}$ pc$^{-3}$]{\label{high_1000_f}{\includegraphics[scale=0.275]{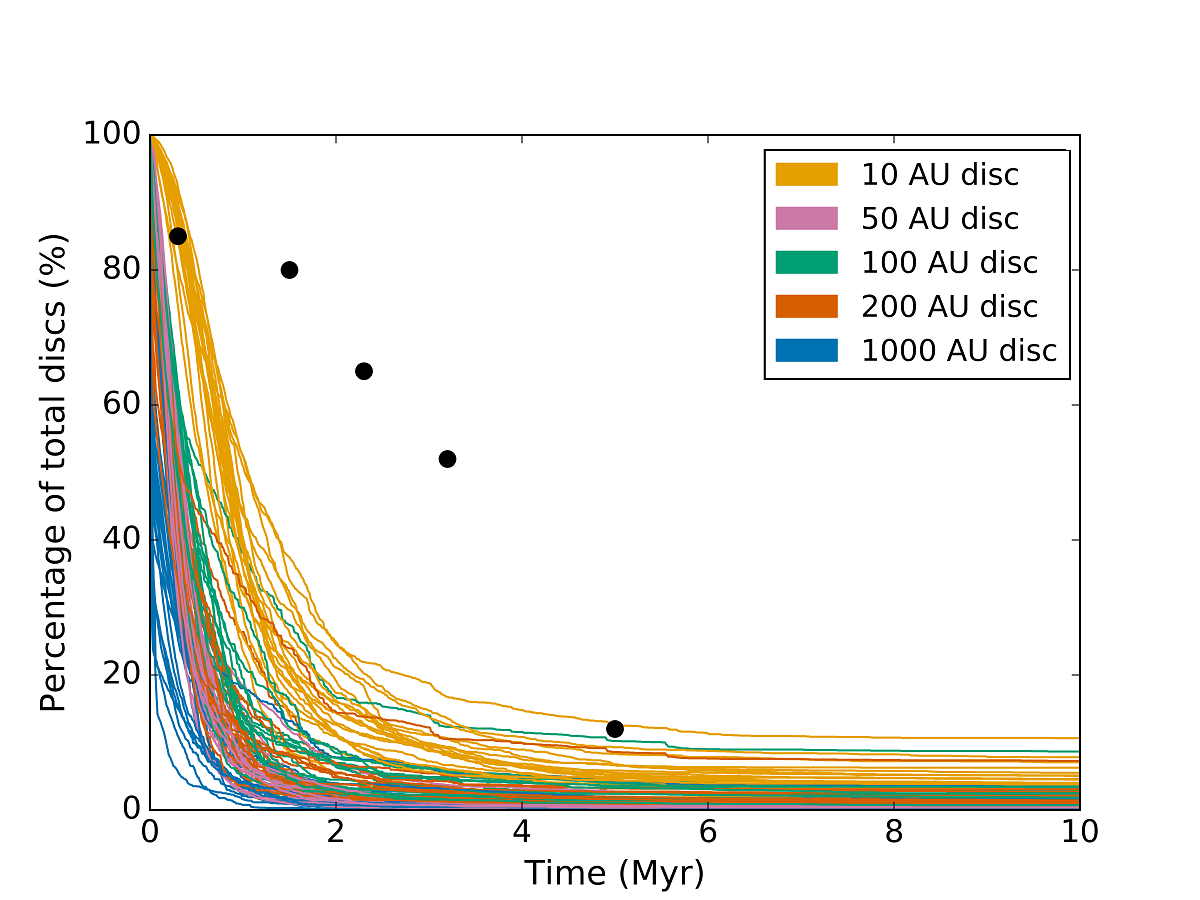}}}
    \caption{The percentage of total remaining discs over time for a star forming region of 1000 M$_{\odot}$ with an initial density of $\sim$10 and 100 M$_{\odot}$ pc$^{-3}$ respectively, a fractal dimension of $D = 2.0$ and a virial ratio of $\alpha_{\rm vir} = 0.3$. The initial disc masses are 1 per cent of the host star mass. Each colour represents a different disc radius. The multiple coloured lines are each a single run of the 20 simulation runs. The black data points are from observational values from \protect\cite{2001ApJ...553L.153H}.}
    \end{center}
\end{figure*}

Fig. ~4(a) shows that on average the time taken for half of the stars within the low density star forming region to lose their 100 AU disc is $0.71$ Myr, less than half of the time taken for discs with 10 per cent of the mass of their stellar host. For discs with a radii of 10 AU, the half life is 3.31 Myr.

The timescale for half of the 100 AU discs to dissipate in the moderately dense (100 M$_{\odot}$\,pc$^{-3}$) star forming region (see Fig.~ 4b) is $\sim$ 0.14 Myr. For discs with a radius of 10 AU, the half life is $\sim$ 0.84 Myr. Less than 5 per cent of 10 AU discs survive for more than 3 Myr.

\begin{table*}
	\centering
	\caption{The time taken for half of stars in a star forming region to lose the gas within their 100 AU protoplanetary discs in a 1000 M$_{\odot}$, moderately sub-structured ($D = 2.0$) region  for two different initial densities; 10 and 100 M$_{\odot}$ pc$^{-3}$ and two different masses of disc, 10 per cent and 1 per cent. Three different virial ratios are analysed: $\alpha_{\rm vir} = 0.3$ (sub-virial, or collapsing), $\alpha_{\rm vir} = 0.5$ (virial equilibrium), and $\alpha_{\rm vir} = 0.7$ (super virial, or expanding). The highest and lowest values from the 20 different runs are included.}
	\begin{tabular}{ccccc} % four columns, alignment for each
		\hline
        & \multicolumn{4}{c|}{Half life of cluster protoplanetary discs} \\
        & \multicolumn{2}{c|}{Disc mass = 0.1\,M$_{\star}$} & \multicolumn{2}{c|}{Disc mass = 0.01\,M$_{\star}$} \\
		Disc Radius (AU) & $\rho = $10 M$_{\odot}$ pc$^{-3}$ & $\rho = $100 M$_{\odot}$ pc$^{-3}$ &  $\rho = $10 M$_{\odot}$ pc$^{-3}$ &  $\rho = $100 M$_{\odot}$ pc$^{-3}$ \\
		\hline
		10 & > 50\% remaining &  3.92 $\pm$ $^{5.91}_{1.87}$ Myr & 3.31 $\pm$ $^{5.16}_{0.9}$ Myr & 0.84 $\pm$ $^{0.22}_{0.27}$ Myr \\ \hline
        50 & 3.94 $\pm$ $^{5.96}_{0.82}$ Myr & 1.04 $\pm$ $^{0.40}_{0.24}$ Myr & 1.22 $\pm$ $^{1.18}_{0.35}$ Myr & 0.28 $\pm$ $^{0.07}_{0.09}$ Myr \\ \hline
        100 & 2.60 $\pm$ $^{1.36}_{0.62}$ Myr & 0.67 $\pm$ $^{0.21}_{0.22}$ Myr & 0.71 $\pm$ $^{0.96}_{0.21}$ Myr & 0.14 $\pm$ $^{0.05}_{0.05}$ Myr \\ \hline
        200 & 1.55 $\pm$ $^{1.34}_{0.44}$ Myr & 0.36 $\pm$ $^{0.09}_{0.11}$ Myr & 0.39 $\pm$ $^{0.58}_{0.12}$ Myr & 0.06 $\pm$ $^{0.03}_{0.02}$ Myr \\ \hline
        1000 & 0.37 $\pm$ $^{0.55}_{0.11}$ Myr & 0.06 $\pm$ $^{0.02}_{0.02}$ Myr & 0.15 $\pm$ $^{0.20}_{0.05}$ Myr & 0.02 $\pm$ $^{0.02}_{0.01}$ Myr \\
		\hline
	\end{tabular}
\end{table*}

\begin{table*}
	\centering
	\caption{The average percentage of 100 AU discs remaining after 10 Myr within a moderately substructured ($D = 2.0$) star forming region for two different initial densities, ($10$ and $100$ M$_{\odot}$ pc$^{-3}$), with two different initial disc masses, 10 per cent and 1 per cent the mass of the host star. The bulk motion  (virial ratio) of the star forming region is varied, from collapsing (sub-virial, $\alpha_{\rm vir} = 0.3$) to expanding (super-virial, $\alpha_{\rm vir} = 0.7$). The highest and lowest values from the 20 different runs are included.}
	\begin{tabular}{ccccc} % four columns, alignment for each
		\hline
         & \multicolumn{4}{c|}{Percentage of discs remaining after 10 Myr} \\
        & \multicolumn{2}{c|}{Disc mass = 0.1\,M$_{\star}$} & \multicolumn{2}{c|}{Disc mass = 0.01\,M$_{\star}$} \\
		Disc Radius (AU) & $\rho = $10 M$_{\odot}$ pc$^{-3}$ & $\rho = $100 M$_{\odot}$ pc$^{-3}$ & $\rho = $10 M$_{\odot}$ pc$^{-3}$ & $\rho = $100 M$_{\odot}$ pc$^{-3}$\\
		\hline
		10 & 77.29 $\pm$ $^{12.79}_{4.71}$ \% & 39.49 $\pm$ $^{14.34}_{14.17}$ \% & 28.81 $\pm$ $^{19.13}_{9.09}$ \% & 3.99 $\pm$ $^{6.65}_{2.18}$ \% \\ \hline
        50 & 34.44 $\pm$ $^{18.08}_{7.57}$ \% & 6.73 $\pm$ $^{5.97}_{3.62}$ \% & 4.79 $\pm$ $^{9.34}_{1.43}$ \% & 0.40 $\pm$ $^{1.53}_{0.23}$ \% \\ \hline
        100 & 17.75 $\pm$ $^{17.03}_{4.04}$ \% & 2.27 $\pm$ $^{4.04}_{0.92}$ \% & 2.27 $\pm$ $^{3.91}_{1.13}$ \% & 0.08 $\pm$ $^{0.59}_{0.08}$ \% \\ \hline
        200 & 7.12 $\pm$ $^{10.79}_{1.82}$ \% & 0.74 $\pm$ $^{1.95}_{0.36}$ \% & 1.09 $\pm$ $^{1.81}_{14.17}$ \% & 0.00 $\pm$ $^{0.38}_{0.00}$ \% \\ \hline
		1000 & 1.09 $\pm$ $^{1.73}_{0.9}$ \% & 0.0 $\pm$ $^{0.38}_{0.00}$ \% & 0.29 $\pm$ $^{0.51}_{0.29}$ \% & 0.00 $\pm$ $^{0.04}_{0.00}$ \% \\
		\hline
	\end{tabular}
\end{table*}

\subsection{Mass of star-forming regions}
We also ran simulations for two different low mass star-forming regions (100 M$_{\odot}$) with an initial density of $\sim$ 100 M$_{\odot}$ pc$^{-3}$, which were sub-virial ($\alpha_{vir}$ = 0.3) and substructured ($D$ = 1.6). These low-mass regions contain one (38\,M$_\odot$) or two (42\,M$_\odot$ and 23\,M$_\odot$) massive stars, which represents an unusual sampling of the IMF \citep{2007MNRAS.380.1271P}, but is observed in nature \citep[e.g. $\gamma^2$ Vel,][]{2014A&A...563A..94J}. Our expectation from randomly sampling the IMF is that 10\,per cent of all star-forming regions can host a massive star, and 1\,per cent of regions will host two massive stars. We note that the lack of massive star(s) in \emph{any} star-forming region would preclude disc destruction from photoevaporation, though as discussed in Section 2 it is unclear which type of star-forming region (in terms of total mass, $M_{\rm cl}$) contributes the most (planet hosting) stars to the Galactic field.

\begin{figure}
	\includegraphics[width=\columnwidth, scale = 5]{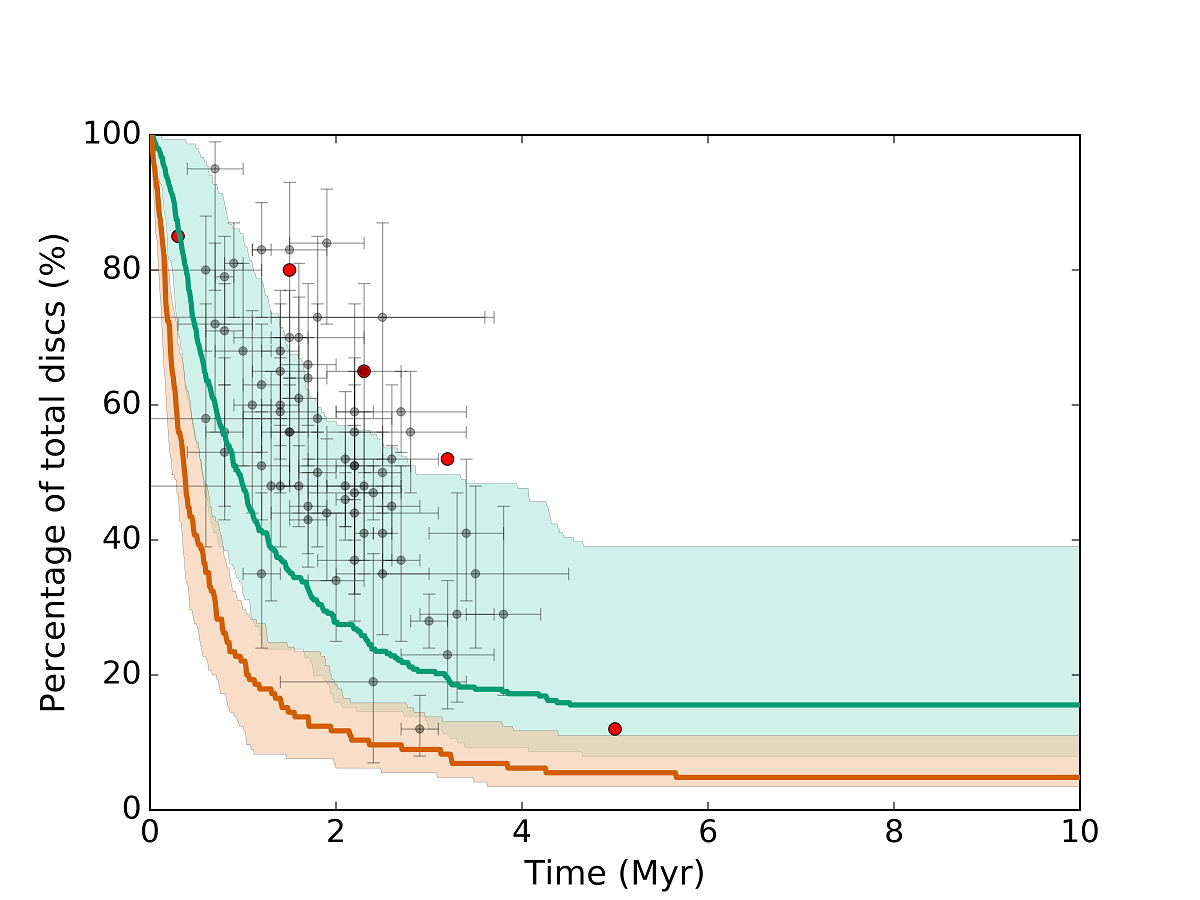}
    \caption{The median percentage of protoplanetary discs (100 AU) remaining with time for two 100 M$_{\odot}$ clusters with initial densities of 100 M$_{\odot}$ pc$^{-3}$ but different numbers of massive stars. The green line shows values for a cluster with 1 massive star (> 15 M$_{\odot}$) and the orange a cluster with 2 massive stars. The red data points are observational values from \protect\cite{2001ApJ...553L.153H}. The grey data points are from \protect\cite{2018MNRAS.477.5191R} using ages from the stellar model in \protect\cite{2000A&A...358..593S}. The colored shaded regions show the complete range of values from the 20 runs for each set of the different clusters.}
    \label{fig:clus_hist}
\end{figure}

In both of these low mass regions, half of discs with radii of 100 AU dissipated before $\sim$ 1 Myr (Fig.~\ref{fig:clus_hist}). The time taken for half of the discs to be destroyed in a region with one massive star is 0.95 Myr. This time is reduced to 0.37 Myr for the cluster with 2 massive stars.

At the end of the 10 Myr simulation, 15.5 per cent of discs within the region with one massive star are surviving. Within the region containing 2 massive stars, less than 5 per cent of discs are remaining, double the number of discs remaining in higher mass regions with the same initial conditions. 

\section{Discussion}
\label{discussion}
The initial conditions of a star-forming region will affect the rate at which protoplanetary discs are photoevaporated due to the radiation from nearby massive stars. The initial substructure and net bulk motion of a star-forming region impacts the rate of disc dispersal. 

\subsection{Changing the initial conditions of star-forming regions}
In our low-density simulations, highly substructured ($D = 1.6$) regions disperse half of the protoplanetary discs within 1.51 Myr, more than twice as fast as smooth ($D = 3.0$) regions. In simulations with a Plummer sphere distribution, more than 30 per cent of discs remain at the end of the 10 Myr simulation, almost double that of discs within highly substructured clusters. 

At these low densities, the degree of substructure matters because a more substructured star-forming region is further from dynamical equilibrium than a smooth region. When this occurs, a low-mass star is likely to have more close encounters with a massive ionising star than in a smooth region. 

In moderately dense initial conditions (100 M$_{\odot}$ pc$^{-3}$), the difference in the fraction of discs that are photoevaporated between different initial spatial distributions decreases greatly, although regions with a Plummer sphere distribution retain more of their discs than regions with initial substructure. However, the average of all runs indicates that the amount of initial substructure has little effect on the survival rates of discs at these densities and fewer than 50 per cent of discs remain after 1 Myr. 

The effect of changing the net bulk motion of the star-forming region has a similar impact on the rate of disc dispersal as the initial substructure has. For low density regions (10 M$_{\odot}$ pc$^{-3}$), the difference between the amount of discs surviving within a collapsing and an expanding star forming region is $\sim$ 15 per cent, with the collapsing regions enabling more photoevaporation than in expanding regions. Again, approximately double the number of discs remain in expanding clusters than in collapsing clusters. In moderately dense clusters it is similar, with the difference being $\sim$ 3 per cent.

For low mass star-forming regions (100 M$_{\odot}$), disc dispersal rates are similar to those of in higher mass regions. Whilst the UV field strength can vary due to different realisations of the IMF \citep{2008ApJ...675.1361F}, these low mass regions show that the mere  presence of a high mass star (> 15\,M$_{\odot}$) will cause disc lifetimes to be shortened dramatically.

Our simulations are set up to mimic the observations of star formation in filaments, where the pre-stellar cores have subvirial motion \citep{1981MNRAS.194..809L, 2015ApJ...799..136F}. The local velocity dispersion is therefore always subvirial to some degree, and because mass-loss due to photoevaporation is so fast (eqns.~2~and~3), most of the photoevaporation occurs during the substructured phase of a star-forming region.

Previous studies investigating the effects of external photoevaporation on disc dispersal rates assumed smooth and centrally concentrated spatial distributions \citep{2001MNRAS.325..449S,2004ApJ...611..360A,2006ApJ...641..504A,2018MNRAS.478.2700W}, replicating environments like the present-day conditions of the ONC. However, using the present-day spatial and kinematic distributions to model star clusters may not accurately replicate the dynamical history of the star-forming region from which the cluster formed \citep{2014MNRAS.438..620P}. 

We cannot provide a direct comparison with the work by \citet{2008ApJ...675.1361F} and \citet{2018MNRAS.478.2700W} because these authors assume initially smooth initial conditions, and we have shown that the severity of photoevaporation depends on the degree of initial substructure, as well as the initial positions of the most massive stars. We have distributed the massive stars randomly within the substructure, and after dynamical evolution these massive stars migrate towards the centre of the star cluster as it forms \citep{2010MNRAS.407.1098A,2014MNRAS.438..620P,2016MNRAS.459L.119P}.

Furthermore, we have used the photoevaporation prescription from \citet{2001MNRAS.325..449S}, rather than determine the photoevaporation rate from the EUV/FUV fluxes as a function of the flux in the interstellar medium \citep[the so-called $G_0$ value, $1.6 \times 10^{-3}$\,erg\,s$^{-1}$\,cm$^{-2}$][]{1968BAN....19..421H}. In comparison with  \citet{2001MNRAS.325..449S}, we find that discs are destroyed earlier in ONC-type regions because the initial densities are higher \citep[in line with current observations,][]{2014MNRAS.445.4037P}, and the star-forming regions are substructured \citep{2004MNRAS.348..589C}.

An initially highly substructured star-forming region can become smooth and centrally concentrated within a few Myr due to a combination of violent and two body relaxation. Protoplanetary discs in these highly substructured environments will be photoevaporated at faster rates than discs within initially smooth regions. Even though they will both appear smooth within a few Myr, the percentage of discs remaining, and possibly the population of planets within the regions, will vary greatly.

The initial density of the cluster has the largest effect on the disc dispersal rate due to external photoevaporation. The `moderately-dense' clusters reflect the likely initial densities of many star-forming regions \citep{2014MNRAS.445.4037P}. However, we find that these `moderately-dense' environments are very destructive for protoplanetary discs and evaporate discs at rates faster than suggested by observations (compare the black points in Figs.~1~and~2 with our simulated data). Our results suggest that protoplanetary discs (or at least their gas content) would always be significantly depleted in moderately dense (100 M$_{\odot}$ pc$^{-3}$) star-forming regions, if those regions contain massive stars. %The time taken for half of the stars in the moderately dense regions to have the gaseous component of their 100 AU discs completely photoevaporated is $0.67$ Myr. For 10 AU discs this is extended to $\sim$ 4 Myr. For 100\,AU discs within the lower density regions, the half-life timescale of the discs is 1.67 Myr.

\cite{2001ApJ...553L.153H} finds that the fraction of disc-hosting stars in young star-forming regions falls to 50\,per cent after $\sim$ 3 Myr whereas \cite{2018MNRAS.477.5191R} find that after only $\sim$ 2 Myr half of the discs remain in their observed regions. In comparison, more than half of the discs in our simulations that are within dense environments are  destroyed within $\sim$1 Myr. One interesting data point in the \cite{2001ApJ...553L.153H} sample is the ONC. With an age of $\sim$1 Myr, the centre of the ONC contains 4 massive stars and a density of $\sim$400 M$_{\odot}$ pc$^{-3}$ \citep{1998ApJ...492..540H}. Observations of this part of the ONC suggest that $\sim$80 -- 85 per cent of stars within the cluster are surrounded by bright ionization fronts, interpreted to be discs, with radii of $\sim$1000 AU \citep{2000AJ....119.2919B,2000AJ....120.3162L}. 

The age of our simulated regions, where 80 per cent of stars still possess a 100 AU disc with some mass, is 0.48 Myr -- likely to be less than half the age of the ONC. This suggests that the massive stars within the ONC should have destroyed the majority of 100 AU discs. From similar arguments, \citet{2007MNRAS.376.1350C} concluded that the possible discs in the ONC with radii >10\,AU are likely to be merely ionisation fronts, containing little mass. Our simulations with different initial disc radii show that the radius of the disc will greatly affect the rate at which it is photoevaporated \citep[see also][]{2007MNRAS.376.1350C} due to the dependence on disc radii within the FUV and EUV photoevaporation prescriptions.

Recent surveys suggest that most stars in the Galactic field host planets, and many of these are gas or ice giants \citep{2011arXiv1109.2497M}. This implies that the majority of planet forming discs were able to survive a significant amount of time in their birth environment. Our simulations suggest that this is only possible in low-density regions that contain no photoionising sources (i.e.\,massive stars). Therefore, (giant) planet formation must occur on very rapid timescales (<1\,Myr), or stars that host giant planets must have formed in very benign environments. %However, when protoplanetary discs are simulated in moderately dense clusters that reflect observations of star forming regions \citep{2016MNRAS.457.3593F}, external radiation from massive stars resulted in large amounts of disc mass loss in a very short time frame and is potentially prohibitive to planet formation.

Many observed protoplanetary discs are located in low-mass, low-density star-forming regions \citep{2013ApJ...771..129A,2018arXiv180305923A} and would be unaffected by external photoevaporation. However, many star-forming regions are typically moderate-density ($\sim$100 M$_{\odot}$ pc$^{-3}$) environments \citep{2014MNRAS.445.4037P} and our results suggest that the majority of protoplanetary discs in star-forming regions with these densities do not survive for long enough periods of time to form giant planets. 

%\textbf{External photoevaporation in these simulations predominantly come from FUV radiation. As the photoevaporation prescription for FUV used does not include information about the mass of the star, there for the mass of the stars within the clusters matters only when considering the effects of EUV, which are minimal.}

%\subsection{Comparison with previous work}

\subsection{Caveats}

There are several caveats to our results, which we discuss below. 

%This lack of a gaseous component in the disc after only a few Myr can greatly effect a variety of processes; from planetary migration to giant planet formation. Our simulations suggest that either gas giants form within very short time scales (> 1 Myr), or dense clusters prohibit the formation of giant planets. 

The effects of external EUV radiation on protoplanetary discs can be reduced when thick winds are present, caused by FUV heating of the disc \citep{2014prpl.conf..475A}. However, the majority of the disc mass loss occurs due to FUV radiation. We repeated our analysis without EUV photoevaporation and find the disc dispersal rates to be similar.

It is possible that we are overestimating the amount of photoevaporation from massive stars. However, recent research suggests that the prescriptions used here are actually underestimating the amount of FUV radiation that discs receive \citep{2016MNRAS.457.3593F,2018MNRAS.475.5460H}. As FUV is the dominant source of external photoevaporation, the protoplanetary discs in our simulations could dissipate on even shorter timescales. 

%{\bf look at graph} We also investigated the effect of changing the EUV flux parameter ($\Phi$) by replacing it with values from other papers (\citep[e.g.][]{2001MNRAS.325..449S}). The rates at which discs were completely photoevaporated did not vary enough to significantly affect the rates of photoevaporation. 

Star formation is an inherently inefficient process, with typically only $\sim$30\,per cent of the mass of a giant molecular cloud converted into stars. Young star-forming regions are observed to contain a large amount of dust and gas, which could shield the proplyds from significant photoevaporation. At these early stages the stellar density within the substructure is highest, and is therefore when the largest percentage of stars are in closest proximity to the massive stars. However, hydrodynamical simulations of star-forming regions show that massive stars blow large ($\sim$pc-scale) cavities within the gas on short time scales \citep{2013MNRAS.430..234D}, and so low-mass disc-hosting stars that would be affected by EUV/FUV radiation will likely reside in the cavities blown out by the massive stars. If the gas and dust could shield the disc, this would protect them for a very short period of time \citep{2015ApJ...804...29G}.  Whether this grace period would be long enough to allow gas-rich giant planets to form is uncertain.

Given that most star-forming regions have stellar densities above a few M$_\odot$\,pc$^{-3}$ \citep{2010MNRAS.409L..54B}, external photoevaporation will detrimentally affect protoplanetary discs in any star-forming region that contains massive stars. This implies that star-forming regions that do not contain massive stars are more likely to form giant planets, but we note that massive stars appear necessary in order to deliver short-lived radioisotopes to the young Solar System \citep{2018PrPNP.102....1L}. The number of massive stars in a star-forming region appears to only be limited by the mass of the star-forming cloud \citep{2007MNRAS.380.1271P}, but this also means that low-mass star forming regions ($<10^4$M$_\odot$) stochastically sample the IMF, meaning that our simulations cannot be described as `typical' star-forming regions.

%The IMF of the star forming regions will greatly affect the rate that discs are photoevaporated. Star forming regions with no massive stars will, of course, not photoevaporate protoplanetary discs, however \cite{2017MNRAS.464.4318N} show that for clusters of all masses that there is a non minimal possibility of even low mass (100 M$_{\odot}$) clusters containing massive stars. Observational counter parts of these low mass clusters with massive stars exist, as mentioned previously. In these lower mass clusters which contain massive stars, discs are photoevaporated at similar rates to discs within higher mass cluster with the same initial conditions. \textbf{I'm kinda lost here.} The probability of stars being born in environments that will photoevaporate protoplanetary discs increases with the mass of the star forming region the star is born into. As a large fraction of stars form in low mass star forming regions \citep{}, the total number of stars affected by external photoevaporation is...
 
Quantifying disc dispersal is further complicated by how difficult it is to determine the ages of young stars, especially before 1 Myr \citep{2000A&A...358..593S}. We use the stellar ages from the \cite{2000A&A...358..593S} model. However, models of pre-main sequence stellar evolution calculate different ages depending on the physics that is implemented. Of the three models presented in \cite{2018MNRAS.477.5191R} we use the ages from \cite{2000A&A...358..593S} so that we are comparing the lower end of cluster ages to our simulations. The average stellar age calculated for the clusters in \cite{2018MNRAS.477.5191R} is significantly shorter than in more recent models from  \citet{2016A&A...593A..99F}. By using these lower age limits, we more than halve the possible average life times of the discs within the observed clusters.

Furthermore, recent work by \citet{2013MNRAS.434..806B} suggests that the ages of pre-main sequence stars may be underestimated by a factor of two, meaning that the observed discs \citep[e.g.][]{2001ApJ...553L.153H} could be a factor of two older. This would make it even more surprising that discs would remain around low-mass stars, if those stars form in regions containing massive stars.

There is also the question of how quickly the photoionising massive stars form. In the competitive accretion models  \citep{2001MNRAS.323..785B}, massive stars gradually gain in mass over $\sim$ 1 Myr \citep{2010ApJ...709...27W}, suggesting high-mass stars form later than low-mass stars \citep{2014prpl.conf..149T}, which would in turn decrease the amount of time low-mass stars spend near the photoionising sources \citep{2012MNRAS.424..377D, 2014MNRAS.442..694D}. In our simulations all stars form simultaneously, and therefore the disc-hosting low-mass stars do not have this grace period, which would increase disc lifetimes.

 %Previous studies, such as \cite{2000A&A...362..968A}, state that clusters richer than the ONC would severely hinder giant planet formation and that the formation of giant planets in such clusters would depend on the time difference between low and high mass star formation. 

The growth of planetesimals into planets can be greatly accelerated by the accretion of cm-scale pebbles. \cite{2017AREPS..45..359J} show that once a 10$^{-2}\,M_{\earth}$ planetesimal has formed it can grow to Jupiter mass in 1 Myr when starting as far out as about 15 AU. An initial phase of accreting pebbles forms a 10$\,M_{\earth}$ core in about 0.8\,Myr, which then undergoes runaway gas accretion to reach Jupiter mass. Such processes potentially allow close-in giant planets to be formed even in the relatively hostile conditions that we consider here.

However, photoevaporation by the central star can cause large amounts of mass loss in the inner disc, potentially affecting giant planet formation \citep{2014prpl.conf..475A}. Grain size also has a significant effect on disc dispersal rates. Mass loss occurs much more quickly when grain growth has occurred because the FUV radiation can penetrate deeper into the disc \citep{2016MNRAS.457.3593F}.

Discs that can survive in moderately dense environments have small radii (10 -- 50 AU). This is because of the disc radius dependency in the external photoevaporation prescriptions. Internal UV radiation can cause significant mass loss and erosion of the disc within short time scales \citep[1 Myr,][]{2008ApJ...683..287G}. The timescale for internal disc dispersal is very short ($10^{5}$ yr), with a UV switch being triggered due to the slowing of accretion onto the inner 10 AU of the disc \citep{2001MNRAS.328..485C}, also calling into question the survivability of small discs. 

Our disc radii are fixed, but in reality disc radii change with time, often in an inside-out fashion where the initial radius is small (and not as susceptible to photoevaporation) compared to later in the disc's life. We include several different disc radii to help visualise what happens for different disc initial conditions, but we cannot model the full viscous evolution in our post-processing analysis.

In our simulations, we have elected to keep the stellar IMF constant across different realisations of the spatial and kinematic initial conditions of our star-forming regions. The reasons for this are two-fold. First, we wish to isolate the possible effects of stochastic dynamical evolution \citep{2010MNRAS.407.1098A,2012MNRAS.424..272P,2014MNRAS.438..620P}, which could lead to different photoevaporation rates even if the ionising flux from massive stars were kept constant. The uncertainties shown by the shaded regions in Figs.~1,~2,~and~5 show this stochasticity for the same initial conditions. Second, the photoevaporation prescriptions we adopt \citep[following][]{2001MNRAS.325..449S} are actually quite insensitive to the mass of the most massive stars (but rather depend on whether the massive stars are present or not). 

However, \citet{2008ApJ...675.1361F} show that the FUV and EUV fluxes can vary if the stellar IMF is extremely top-heavy and contains more massive stars than expected on average. In a forthcoming paper we will calculate the EUV and FUV fluxes in our substructured star-forming regions and use recent the FRIED models of disc photoevaporation from \citet{2018MNRAS.481..452H} to determine mass-loss based on these models, and whether it depends strongly on stochastic sampling of the stellar IMF.

Similarly, in some of our simulations the massive stars are ejected early on, which is a common occurrence in simulations of dense star-forming regions \citep[][Schoettler et al., in prep.]{2010MNRAS.407.1098A,2016A&A...590A.107O}. We will also quantify the effects of these ejections on the fraction of surviving protoplanetary discs in an upcoming paper.

The majority of discs observed with ALMA have been located in low-mass, low-density star-forming regions. Current observations suggest that the majority of stars form in moderately dense ($\sim$100 M$_{\odot}$ pc$^{-3}$) environments \citep{2014MNRAS.445.4037P}. However, the majority of protoplanetary discs in clusters with these densities do not survive for long enough periods of time to form planets, as planet formation is thought to take place over a few million years \citep{1996Icar..124...62P}. The fact that the majority of stars have planetary systems around them poses important questions as a result of the discrepancies that seemingly arise. This may indicate that the majority of stars form in low mass clusters where there are few to no high mass stars.

We adopt initial disc masses that are 10 per cent the mass of the host star, which is likely to be a large overestimate. When looking at more realistic values (1 per cent), discs are destroyed on even shorter timescales. However, it should be noted that accretion and internal photoevaporation will have much larger effects on disc mass evolution for these lower mass discs. % The rate at which these discs are destroyed may pose some fundamental questions for our understanding of planet and star formation.

%Observations of star forming regions that contain both high and low density environments are necessary to resolving some of the contradictory points raised throughout the field of planet formation. By studying of disc fractions in complex star forming regions, such as Corina, we can begin to disentangle the effects that the environment plays on protoplanetary disc formation and evolution.

\section{Conclusions}
\label{conclusion}
We have calculated the mass loss due to external photoevaporation of protoplanetary discs in $N$-body simulations of the evolution of star-forming regions. We ran a suite of simulations where we vary the initial spatial structure, bulk motion and initial density of the regions. We compared our simulations that more closely represent observed star-forming regions (subvirial, substructured) with those of primordially mass segregated, spherical clusters, similar to those used in previous studies of external photoevaporation.

The parameter that most affects rates of disc dispersal is the initial density of the star-forming region. The majority of protoplanetary discs within simulated regions that mimic the conditions in nearby star-forming regions are dispersed due to external photoevaporation within very short time scales. In moderately dense ($\sim$100 M$_{\odot}$ pc$^{-3}$) star-forming regions which have moderate levels of substructure ($D = 2.0$) and are collapsing ($\alpha_{\rm vir} = 0.3$), we find the time taken for half of 100 AU discs to dissipate is significantly shorter (3 times less) than suggested in observational studies \citep{2001ApJ...553L.153H}. Lower density clusters ($\sim$10 M$_{\odot}$ pc$^{-3}$) allow discs to survive long enough to match observations of disc lifetimes, although the half-life of 100 AU discs is still less than that found by \cite{2001ApJ...553L.153H}.

The initial spatial distribution of the star-forming region also affects the rate of protoplanetary disc dispersal due to external photoevaporation. The degree to which initial substructure affects disc dispersal rates depends on the initial density. In moderately dense ($\sim$100 M$_{\odot}$ pc$^{-3}$) regions the effects are washed out, but in lower-density regions ($\sim$10 M$_{\odot}$ pc$^{-3}$) we find that the more fractal and clumpy a star-forming region is, the higher the rate of disc dispersal. This is due to violent relaxation and the rapid increase in density (sometimes up to an order of a magnitude) of the star forming region within a short amount of time. As most star forming regions appear to have a high degree of substructure, it is important for future studies of disc dispersal to take the initial conditions into consideration due to external photoevaporation in dense environments.

The virial ratio of the star forming region affects the rate of disc dispersal in a similar way to substructure. Regions that have a low initial density and are collapsing photoevaporate more discs on average than clusters which are expanding. The effects of varying the initial net bulk motion in moderately density clusters is negligible. 

The majority of observed stars in the Galactic field host planetary systems, implying their protoplanetary discs survived long enough for formation to take place. There are three possible scenarios to resolve this apparent tension between observations and our simulations: %Our simulations of protoplanetary discs in cluster environments similar to that of the ONC showed that, when subjected to external photoevaporation from neighboring massive stars, the majority are destroyed within 3 Myr. The time taken for half of stars to lose their disc was $\sim$ 1 Myr, half the time for observed discs. Approximately $\sim$ 80 percent of stars within the ONC are disc bearing at $\sim$ 1 Myrs. In comparison, the age at which 80 percent of stars in our simulated cluster still possess 100 Au discs with some amount of mass is 0.48 Myr.

%The discrepancies between our simulations and observations point towards several possibilities: 

i) The majority of planets may not form in moderately dense star-forming regions ($\sim$100 M$_{\odot}$ pc$^{-3}$); rather, they would form in low density regions with no photoionising massive stars present. Many of the protoplanetary discs have been in observed in these low-density ambient environments \citep{2018arXiv180305923A}, but significant numbers of protoplanetary discs (or at least their remnants) have been observed in dense, hostile regions like the ONC \citep{1996AJ....111.1977M}.

ii) If some planets do form in dense, clustered environments containing massive stars (such as the ONC), then this suggests that giant planet formation must happen on very short time scales (less than 1 -- 2 Myr), or be confined to discs with radii significantly smaller than the orbit of Neptune in our Solar System. \cite{2017AREPS..45..359J} show that giant planet formation can occur on these timescales once large enough planetesimals have formed. However, internal photoevaporation processes can deplete the inner disc and set limits on the formation time of giant planets \citep{2014prpl.conf..475A}.

iii) The current calculations of mass-loss in discs due to external photoevaporation are severely overestimating the detrimental effects of EUV and FUV radiation. However, recent research \citep{2016MNRAS.457.3593F,2018MNRAS.475.5460H} suggests that photoevaporative mass-loss rates caused by FUV radiation may be underestimated, and our calculations also underestimated the effects as we adopt conservatively high initial disc masses.

%RJP OUT

\section*{Acknowledgements}
We thank A. R. Williams for useful discussions and S. Habergham-Mawson for comments. RBN is partially supported by an STFC studentship. RJP acknowledges support from the Royal Society in the form of a Dorothy Hodgkin Fellowship. RC and MBD are supported by grant 2014.0017, ``IMPACT", from the Knut and Alice Wallenberg Foundation.

%%%%%%%%%%%%%%%%%%%%%%%%%%%%%%%%%%%%%%%%%%%%%%%%%%



\bibliographystyle{mnras}
\bibliography{bib} % if your bibtex file is called example.bib

%%%%%%%%%%%%%%%%%%%%%%%%%%%%%%%%%%%%%%%%%%%%%%%%%%

%%%%%%%%%%%%%%%%% APPENDICES %%%%%%%%%%%%%%%%%%%%%

\appendix

%%%%%%%%%%%%%%%%%%%%%%%%%%%%%%%%%%%%%%%%%%%%%%%%%%

% Don't change these lines
\bsp	% typesetting comment
\label{lastpage}
\end{document}